\documentclass{article} 

\usepackage[final]{neurips_2025}

\usepackage{amsmath,amsfonts,bm}

\def\eqref#1{equation~\ref{#1}}

\def\1{\bm{1}}

\def\rx{{\textnormal{x}}}

\def\rvh{{\mathbf{h}}}

\def\rvw{{\mathbf{w}}}
\def\rvx{{\mathbf{x}}}
\def\rvy{{\mathbf{y}}}

\def\mI{{\bm{I}}}

\def\mQ{{\bm{Q}}}

\DeclareMathAlphabet{\mathsfit}{\encodingdefault}{\sfdefault}{m}{sl}
\SetMathAlphabet{\mathsfit}{bold}{\encodingdefault}{\sfdefault}{bx}{n}

\usepackage{url}
\usepackage{graphicx}
\usepackage{booktabs}
\usepackage[utf8]{inputenc} 
\usepackage[T1]{fontenc}    
\usepackage{amsfonts}       
\usepackage{nicefrac}       
\usepackage{microtype}      
\usepackage{xcolor}         
\usepackage{algorithm}
\usepackage{algpseudocode}
\usepackage{hyperref}
\usepackage{multirow}

\title{Steering Generative Models with Experimental Data for Protein Fitness Optimization}

\author{
    Jason Yang$^{\dagger}$\\
    Chemistry \& Chemical Engineering \\
    California Institute of Technology \\
  \And
    Wenda Chu$^{\dagger}$\\
    Computing \& Mathematical Sciences \\
    California Institute of Technology \\
  \And
    Daniel Khalil \\
    Computing \& Mathematical Sciences \\
    California Institute of Technology \\
    \And
    Raul Astudillo \\
    Computing \& Mathematical Sciences \\
    California Institute of Technology \\
  \And
    Bruce J. Wittmann \\
    Office of the Chief Scientific Officer \\
    Microsoft Corporation \\
  \And
    Frances H. Arnold \\
    Chemistry \& Chemical Engineering \\
    Biology \& Biological Engineering \\
    California Institute of Technology \\
  \And
    Yisong Yue\thanks{yyue@caltech.edu}\\
    Computing \& Mathematical Sciences \\
    California Institute of Technology \\
}

\begin{document}

\maketitle
\def\thefootnote{$\dagger$}\footnotetext{These authors contributed equally to this work.}\def\thefootnote{\arabic{footnote}}

\begin{abstract}
Protein fitness optimization involves finding a protein sequence that maximizes desired quantitative properties in a combinatorially large design space of possible sequences. Recent advances in steering protein generative models (e.g., diffusion models and language models) with labeled data offer a promising approach. However, most previous studies have optimized surrogate rewards and/or utilized large amounts of labeled data for steering, making it unclear how well existing methods perform and compare to each other in real-world optimization campaigns where fitness is measured through low-throughput wet-lab assays. In this study, we explore fitness optimization using small amounts (hundreds) of labeled sequence-fitness pairs and comprehensively evaluate strategies such as classifier guidance and posterior sampling for guiding generation from different discrete diffusion models of protein sequences. We also demonstrate how guidance can be integrated into adaptive sequence selection akin to Thompson sampling in Bayesian optimization, showing that plug-and-play guidance strategies offer advantages over alternatives such as reinforcement learning with protein language models. Overall, we provide practical insights into how to effectively steer modern generative models for next-generation protein fitness optimization.
\end{abstract}

\section{Introduction}

Proteins, sequences of amino acids, can be optimized for useful properties such as binding affinity, catalytic activity, or stability, numerically quantified as ``fitness.'' However, protein optimization is challenging: the design space of proteins is enormous, as a protein of length $M$ can be constructed in $20^M$ different ways, of which only a negligible fraction are functional \citep{romero_exploring_2009}. Moreover, most  wet-lab assays only provide $10^2-10^3$ fitness labels. Consequently, researchers often rely on directed evolution, an iterative process aiming to incrementally improve protein fitness \citep{packer_methods_2015} through multiple rounds of mutation and experimental screening. In each round, a protein is mutated, the variants’ fitnesses are measured, and the most beneficial variant is selected for the next iteration. However, this approach can be slow, often accumulating only one mutation per round, and inefficient, as it performs a local search limited to closely related protein sequences. 

\begin{figure}[t]
  \centering
\includegraphics[width=\textwidth]{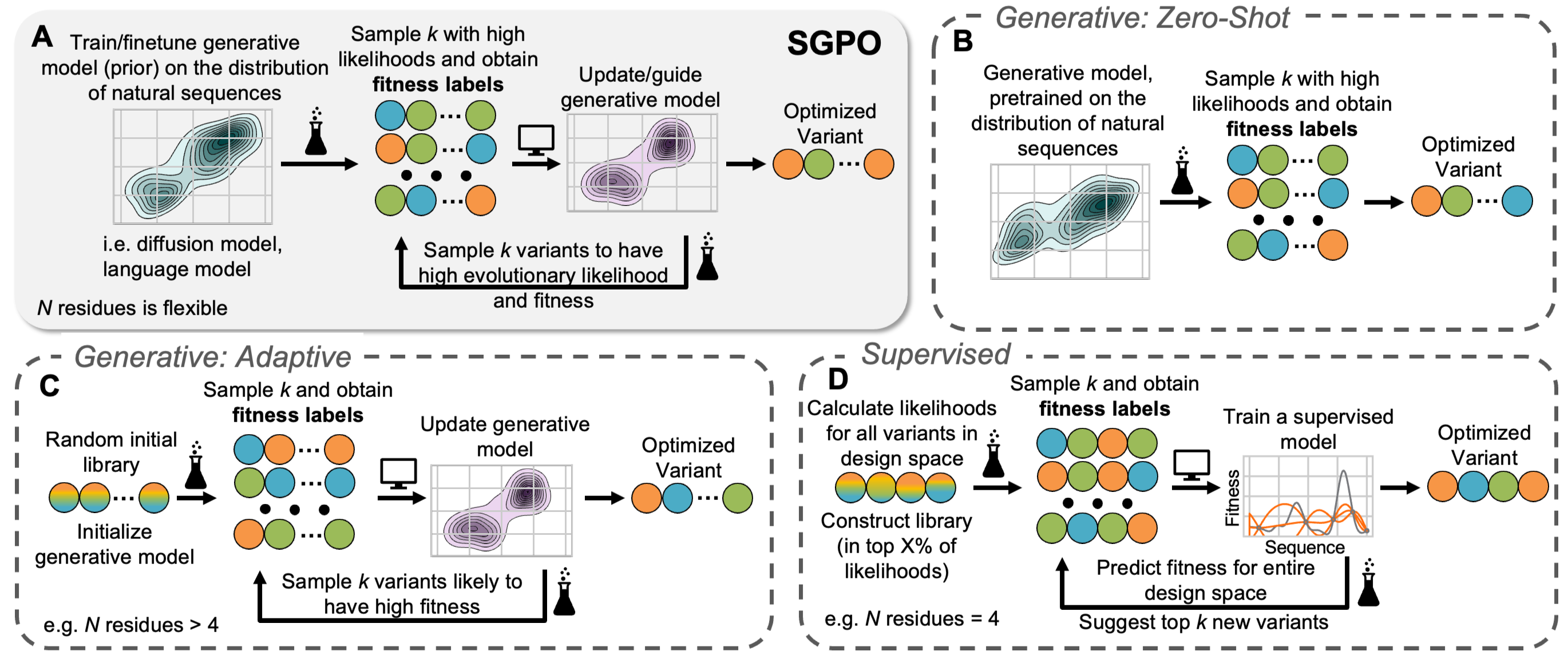}
  \vspace{-0.3in}
  \caption{\textbf{Comparison of steered generation for protein optimization (SGPO) to other ML-assisted workflows for protein engineering.} \textbf{(A)} SGPO involves initializing a generative prior model to sample sequences with high natural likelihoods and steering that model with assay-labeled fitness data. Optimization is difficult because the design space is vast, and the throughput of wet-lab fitness assays (Erlenmeyer flask icon) is low, so adaptive learning across multiple iterations is beneficial. Previous methods have utilized generative models such as \textbf{(B)} fully zero-shot methods that sample highly natural sequences but do not utilize labeled fitness data or \textbf{(C)} those that only utilize labeled fitness. \textbf{(D)} Alternatively, supervised approaches involve enumerating to calculate fitness predictions for all variants in a design space, limiting them to optimizing few residues (i.e., $N < 9$).}
    \vspace{-0.15 in}
  \label{fig:background}
\end{figure}

In recent years, there has been a growing interest in developing machine learning (ML)-assisted methods to optimize protein fitness more efficiently \citep{yang_machine-learning-guided_2019, wittmann_advances_2021, hie_adaptive_2022, yang_opportunities_2024, yang_illuminating_2025}. Many recent studies have focused on generative approaches combining unlabeled and labeled data for protein design. Broadly, these methods achieve \textit{conditional generation} by steering generative priors of natural protein sequences \citep{freschlin_machine_2022} using fitness data, thereby enabling incorporation of the steered models into adaptive optimization cycles \citep{hie_adaptive_2022}. We refer to this class of methods as \textbf{S}teered \textbf{G}eneration for \textbf{P}rotein \textbf{O}ptimization (SGPO). These methods address the individual limitations of previous approaches (Fig. \ref{fig:background}A, Table \ref{table:motivation}). First, SGPO leverages labeled data, which is essential for fitness goals that deviate from natural functions (e.g., engineering enzymes for non-native activities \citep{arnold_directed_2018, yang_active_2025}), unlike zero-shot methods relying solely on  generative priors of natural sequences (\textit{Generative: Zero-Shot}, Fig. \ref{fig:background}B, \citealt{hie_efficient_2023, sumida_improving_2024, fei_advancing_2025, seki_combinatorial_2025, lambert_sequence-based_nodate}). Second, generative priors \citep{wu_protein_2021, hsu_generative_2024} sample sequences with high evolutionary likelihoods and potentially higher fitness, giving these methods a significant advantage over approaches relying exclusively on labeled data \citep{ song_importance_2024, stanton_accelerating_2022, gupta_feedback_2019, brookes_design_2020, jain_biological_2022, kim_improved_2025, angermueller_model-based_2020, hie_adaptive_2022} (\textit{Generative: Adaptive}, Fig. \ref{fig:background}C). Finally, SGPO scales to larger design spaces, unlike most supervised ML-assisted directed evolution (MLDE) approaches, which require enumerating and scoring all variants in the design space (\textit{Supervised}, Fig. \ref{fig:background}D) \citep{wu_machine_2019, wittmann_informed_2021, yang_active_2025, li_evaluation_2024, vornholt_enhanced_2024,jiang_rapid_2024, hsu_learning_2022, ding_machine_2024, hawkins-hooker_likelihood-based_2024, zhao_contrastive_2024, thomas_engineering_2025, sun_accelerating_2025}.

Despite these advantages, SGPO methods still face practical limitations in real-world fitness optimization, particularly across two major classes of approaches: guiding discrete diffusion models \citep{nisonoff_unlocking_2024, stark_dirichlet_2024, klarner_context-guided_2024, gruver_protein_2023, lisanza_multistate_2024, goel_memdlm_2024} and finetuning models such as protein language models (PLMs) through reinforcement learning (RL) \citep{ruffolo_designing_2024, widatalla_aligning_2024, stocco_guiding_nodate, blalock_functional_nodate, wang_proteinzero_2025}. The limitations of prior work are summarized as follows: (1) Few previous studies have explored steering with few ($10^2-10^3$) labeled sequences \citep{lisanza_multistate_2024, stocco_guiding_nodate} for protein optimization based on real fitness data, e.g. activity or fluorescence, rather than computational surrogates \citep{lisanza_multistate_2024, blalock_functional_nodate}. (2) Most studies only evaluate one type of generative prior and steering strategy, so it is unclear how different combinations perform in practice. (3) There is room to incorporate principles from adaptive optimization, such as uncertainty-aware exploration (e.g., Bayesian optimization), which have shown clear benefits in protein engineering \citep{vornholt_enhanced_2024, yang_active_2025}. 

\begin{table}[t]
  \caption{\textbf{SGPO is a general approach for protein fitness optimization that does not face the individual limitations of other strategies.} Namely, SGPO utilizes zero-shot knowledge from the natural distribution of proteins, can be guided by assay-labeled fitness data, and can optimize many residues ($N$) simultaneously. Beyond those listed here, there are many other studies that combine different elements of these approaches.}
  \vspace{-0.2in}
  \centering
  \begin{tabular}{p{1.6cm}  p{1.5cm} p{1.2cm}  p{1.2cm} p{6.4cm} }
    \\
    \toprule
    Approach & Prior Information Used? & Assay Fitness Used? & Scales to large $N$?  & Protein Examples (non-exhaustive) \\
    \midrule
    \textbf{SGPO}    & \checkmark & \checkmark &   \checkmark & \citet{lisanza_multistate_2024, widatalla_aligning_2024, stocco_guiding_nodate, nisonoff_unlocking_2024, brookes_conditioning_nodate, blalock_functional_nodate, goel_memdlm_2024, pmlr-v267-huang25ba} \\
    \midrule
    Generative: Zero-Shot     & \checkmark & $\times$ &   \checkmark & \citet{hie_efficient_2023, sumida_improving_2024, fei_advancing_2025, seki_combinatorial_2025, lambert_sequence-based_nodate}\\
    \midrule
    Generative: Adaptive     & $\times$ & \checkmark &  \checkmark & \citet{song_importance_2024, jain_biological_2022, angermueller_model-based_2020, stanton_accelerating_2022, brookes_design_2020} \\
    \midrule
     Supervised     & \checkmark & \checkmark & $\times$ & \citet{wittmann_informed_2021, ding_machine_2024, hawkins-hooker_likelihood-based_2024, zhao_contrastive_2024, sun_accelerating_2025}\\ 
    \bottomrule
  \end{tabular}
  \vspace{-0.2in}
  \label{table:motivation}
\end{table}

In this study, we aim to understand the best practices for integrating SGPO into real-world engineering workflows. We focus here on modern generative models (i.e., discrete diffusion, language models) but acknowledge that other related methods are relevant, such as those based on variational autoencoders \citep{brookes_conditioning_nodate, torres_generative_2024} and other adaptive search strategies \citep{kirjner2024improvingproteinoptimizationsmoothed, sinai2020adaleadsimplerobustadaptive, pmlr-v162-ren22a}. We explore the following questions: Which steering strategies perform best, and with which types of models? How can we utilize uncertainty to better explore the design space when performing guidance? \textbf{Overall, we make the following key contributions:}
\begin{enumerate}
    \item We motivate SGPO as a useful, general framework and contextualize existing methods for protein optimization under this umbrella.
    \item We comprehensively evaluate design decisions for SGPO, including different generative models for sequences and steering strategies (Fig. \ref{fig:diffusion} \& \ref{fig:landscape}, Section \ref{section:strategies}), offering best practices for protein  optimization with few fitness labels.
    \item We introduce ideas from adaptive optimization into SGPO by proposing a method that ensembles multiple plug-and-play fitness predictors and leverages their predictive uncertainty to enable more efficient exploration.
    \item We are the first to adapt \textit{decoupled annealing posterior sampling} \citep{zhang_improving_2024} for SGPO, and this type of plug-and-play guidance has the strongest performance overall.
\end{enumerate}

On the TrpB, CreiLOV, and GB1 protein fitness datasets, we find that SGPO methods can consistently identify high-fitness protein variants. In particular, our results highlight the advantages of plug-and-play guidance with diffusion models over finetuned language models—offering greater steerability and lower computational cost. To support future research and real-world adoption, our extensive, user-friendly code is available at \url{ https://github.com/jsunn-y/SGPO}.

\section{Related work}
\label{section:strategies}

\begin{figure}[h]
  \centering
\includegraphics[width=\textwidth]{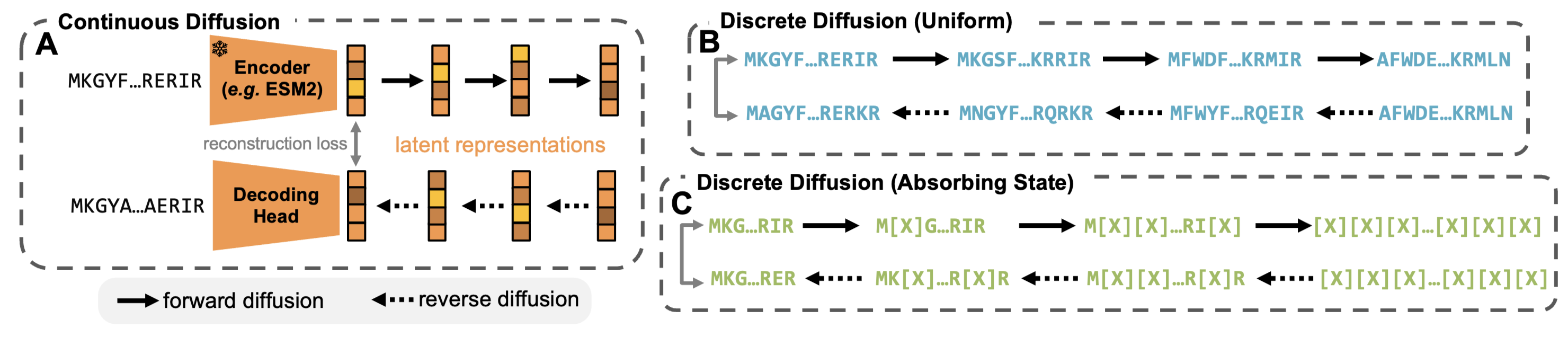}
  \vspace{-0.25in}
  \caption{\textbf{Overview of different approaches to train diffusion models over discrete state spaces.} During inference, a noised latent representation or sequence is decoded into a reasonable sequence (bottom track for each method). [X] refers to a masked token.}
    \vspace{-0.2in}
  \label{fig:diffusion}
\end{figure}

\paragraph{Generative models for discrete sequences.}
The most widely adopted generative models for natural protein sequences are PLMs, such as autoregressive transformers \citep{nijkamp_progen2_2023} and masked language models \citep{rives_biological_2021}. Increasingly, various diffusion model \citep{ho_denoising_2020} architectures have shown efficacy for modeling discrete data ($\rvx$) \citep{li_survey_2025}, such as protein sequences \citep{alamdari_protein_2023, wang_diffusion_2024}, leveraging many similar learning techniques such as masking or autoregressive decoding \citep{sahoo_simple_2024, lou_discrete_2024, nie_large_2025, shi2025simplifiedgeneralizedmaskeddiffusion} (Fig. \ref{fig:diffusion}). These generative prior models $p(\rvx)$ can be categorized broadly into two types: those that perform diffusion in a continuous latent space \citep{li_diffusion-lm_2022, chen_analog_2023, dieleman_continuous_2022} and those that diffuse directly over discrete space (Fig. \ref{fig:diffusion}). In the protein domain, it has also been shown that latent diffusion over embeddings from PLMs can be more effective \citep{meshchaninov_diffusion_2025, chen_amp-diffusion_2024, torres_generative_2025}. Alternatively, models performing diffusion in discrete space use a transition matrix to update all discrete states in each timestep (D3PM) \citep{austin_structured_2023}, which has later been formulated as continuous-time Markov chains \citep{lou_discrete_2024, campbell_continuous_2022, campbell_generative_2024, schiff_simple_2024}. Two common ways to add noise to discrete sequences are to use uniform noise matrices or absorbing state (masking) matrices (Fig. \ref{fig:diffusion}). These have been followed by simplified frameworks showing some of the highest performance for modeling natural language, such as masked diffusion language models (MDLMs) \citep{sahoo_simple_2024, hoogeboom_autoregressive_2022, shi2025simplifiedgeneralizedmaskeddiffusion} and a variation that uses uniform noise called uniform diffusion language models (UDLMs) \citep{schiff_simple_2024}. We elaborate more on these methods in Section \ref{section:model_methods}.

\begin{figure}[h]
  \centering
\includegraphics[width=\textwidth]{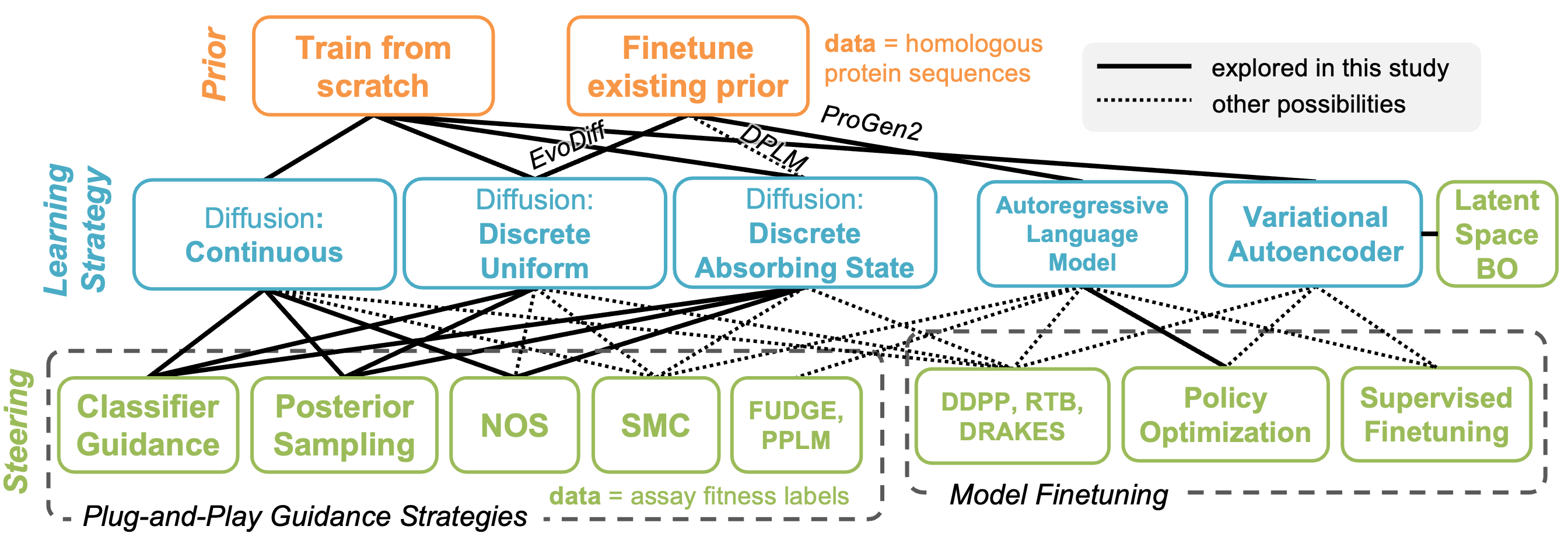}
  \vspace{-0.25in}
  \caption{\textbf{Methods design space for SGPO: a non-exhaustive landscape of generative models for protein \textit{sequences} and methods to steer them with labeled data.} Three major types of diffusion models for sequences include those that perform diffusion over continuous space and those that perform diffusion over discrete space with a uniform or absorbing state (masking) noising process. Various types of guidance strategies are compatible with certain models, in green (NOS: diffusion optimization sampling, SMC: sequential monte carlo, FUDGE: future discriminators for generation, PPLM: plug and play language models, DDPP: discrete denoising posterior prediction, RTB: relative trajectory balance, DPLM: diffusion protein language model, BO: Bayesian optimization). Differently, language models and variational autoencoders can be aligned with labeled data via reinforcement learning such as policy optimization or supervised finetuning.}
    \vspace{-0.2in}
  \label{fig:landscape}
\end{figure}

\paragraph{Plug-and-play guidance strategies.}
An advantage of diffusion models is the ability to perform plug-and-play guidance based on fitness labels ($\rvy$) without finetuning the generative prior model weights, resulting in reduced training costs and potentially strong signal despite having few ($\sim10^2)$ labels.
Guiding a continuous diffusion model often involves skewing the learned score function using gradients from a supervised value function that can predict labels $\rvy$ from data $\rvx$~\citep{chung2023diffusion, zheng2025inversebench,soares_targeted_2025}. These methods are often referred to as posterior sampling, as they aim to sample from the posterior distribution, $p(\rvx|\rvy)$.
Recent works extend this idea to guiding discrete diffusion models. Classifier guidance (CG) \citep{nisonoff_unlocking_2024} skews the rate matrix of the reverse time Markov chain of discrete diffusion models using a time-dependent value function, $p(\rvy|\rvx_t,t)$; variable splitting methods (DAPS) \citep{zhang_improving_2024,chu2025split} use discrete diffusion models as denoisers and only require a value function of clean data, $p(\rvy|\rvx_0)$; diffusion optimized sampling (NOS) \citep{gruver_protein_2023} trains a value function on continuous embeddings of discrete tokens and optimizes the embedding for higher fitness; sequential Monte Carlo methods (SMC)~\citep{li_derivative-free_2024,uehara_reward-guided_2025,wu2024practical, lee_debiasing_2025, singhal_general_2025} evolve multiple particles from a series of distributions to approximate the posterior distribution in limit.
We explain these methods in more detail in Section \ref{section:steering_methods}, along with other variations on the guidance process. In this study, we focus on CG, DAPS, and NOS as guidance techniques (Fig. \ref{fig:landscape}). Future work could also consider guidance techniques for autoregressive language models, such as future discriminators for generation (FUDGE) \citep{yang_fudge_2021}, plug and play language models (PPLM) \citep{dathathri_plug_2020}, and twisted SMC \citep{zhao_probabilistic_2024, amin_bayesian_2024}. Additionally, \citet{xiong_guide_2025} demonstrate how guidance generalizes to masked language models and order-agnostic autoregressive models.

\paragraph{Reinforcement learning via model finetuning.}
We consider RL broadly here as techniques that achieve conditional generation by finetuning generative models with labeled data, thus pushing those models to produce more favorable generations. There are emerging RL techniques applied to discrete diffusion models, including discrete denoising posterior prediction (DDPP) \citep{rector-brooks_steering_2024}, relative trajectory balance (RTB) \citep{venkatraman_amortizing_2025, bartoldson_trajectory_2025, venkatraman_outsourced_2025}, and direct reward backpropagation with gumbel softmax trick (DRAKES) \citep{wang_fine-tuning_2024}. While the above strategies are specific to discrete diffusion models, supervised fine-tuning (SFT) and policy optimization are two important techniques used in RL that can be broadly applied to generative models such as language models (Fig. \ref{fig:landscape}). Policy optimization has generally shown better performance than SFT \citep{stocco_guiding_nodate, blalock_functional_nodate}; in particular, direct preference optimization (DPO) is often used for its algorithmic simplicity and ease of training \citep{rafailov_direct_2023} (details in Section \ref{section:steering_methods}). RL has demonstrated utility for aligning generative models of proteins (language models, inverse folding models, variational autoencoders) with properties like stability \citep{widatalla_aligning_2024, blalock_functional_nodate, stocco_guiding_nodate, lim_scoring-assisted_2025}, but these methods can have high computational costs of finetuning and may require large amounts of labels ($> 10^3$) to effectively steer generations. We include DPO with an autoregressive PLM (finetuned ProGen2 \citep{nijkamp_progen2_2023}) as a baseline.

\paragraph{Adaptive optimization.}
Protein engineering is commonly conducted through adaptive workflows such as directed evolution \cite{packer_methods_2015} or ML-based approaches such as Bayesian optimization \citep{frazier2018tutorial, stanton_accelerating_2022}. These methods follow an iterative loop: labeled data is collected via expensive wet-lab assays, a surrogate model $p(\rvy|\rvx)$ is trained or updated, an acquisition function implied by the surrogate is used to propose new sequences to evaluate, and the cycle repeats \citep{hie_adaptive_2022, vornholt_enhanced_2024, yang_active_2025}. The surrogate model, often a Gaussian process or a deep ensemble, provides uncertainty estimates, which are used by an acquisition function (e.g., expected improvement, Thompson sampling) to balance exploration and exploitation of the design space. In this study, we adapt these ideas to guide diffusion models for protein sequence generation, as described in Section~\ref{section:iterative}. A closely related line of work is latent space Bayesian optimization \citep{maus2022local, stanton_accelerating_2022, gomez-bombarelli_automatic_2018, castro_guided_2022, torres_generative_2024, lee_latent_2025}, which searches for optimal sequences within a latent space—typically learned by an autoencoder, which can implicitly capture a prior on natural protein sequences. In this work, we compare against APEXGo \citep{torres_generative_2024}, a method that performs trust-region Bayesian optimization in the latent space of a variational autoencoder trained over protein sequences. There are also related methods that involve conditional sampling from a prior \citep{brookes_conditioning_nodate}. However, we note that SGPO offers greater flexibility by avoiding reliance on an explicit latent space, which enables the use of modern, more powerful generative models such as diffusion models and protein language models that are not easily accommodated by traditional latent Bayesian optimization pipelines.

\section{Problem setup}

We focus on evaluating methods that fall under SGPO, where the primary downstream task entails starting from a known sequence with some level of fitness for a target objective (i.e. activity, stability, fluorescence, binding, etc.) and identifying a modified sequence with maximized fitness, where real-world fitness can only be measured for $10^2$ to $10^3$ sequences. Our goal is to sample sequences with maximum fitness $\rvy$ from the \textit{generative prior} $p(\rvx)$, which is trained on the multiple sequence alignment (MSA) of homologous protein sequences that are evolutionarily related to a known protein with some level of desired fitness (details in Section \ref{section:model_methods}). This model can be thought of as capturing the distribution of sequences with high likelihood from a given protein family. 

During inference, sequences can be sampled \textit{unconditionally} from $p(\rvx)$, or sampling can be \textit{guided} using a supervised model of the form $p(\rvy|\rvx) \propto \exp(f(\rvx)/\beta)$, where $f(\cdot)$ is a learned fitness predictor—also referred to as the \textit{classifier} or \textit{value function}. This predictor is trained on a small number of labeled sequence-fitness pairs (typically in the hundreds) to reflect practical data limitations. The goal of guided sampling is to generate protein sequences from the posterior distribution, $p(\rvx|\rvy) \propto p(\rvx) \exp(f(\rvx)/\beta)$. We use a computational \textit{oracle} to acquire and evaluate fitness labels $\rvy$, to simulate how fitness would be measured in a real-world campaign.  Details on training and guidance with the value function are provided in Section \ref{section:steering_methods} and Table \ref{table:classifier_training}. As an alternative steering method to guidance, we finetune the generative prior with labeled data using an autoregressive language model (ARLM) and DPO, which serves as a baseline. We further compare to a baseline of latent space Bayesian optimization. The strength of steering is tuned by method-specific hyperparameters.

\section{Results}

\begin{table}[h]
  \caption{\textbf{Summary of datasets used in this work.} Train and test fitness refer to the number of fitness labels used for training and testing the oracle. We focus on TrpB and CreiLOV, with some of the GB1 results moved to the Appendix. While the TrpB dataset has a lot more training labels, it may be more difficult to learn due to relatively high amounts of epistatic effects between residues (non-additivity of mutation effects).}
  \vspace{-0.2in}
  \centering
  \begin{tabular}{p{1.87cm}  p{0.78cm} p{3.3cm}  p{0.8cm} p{0.6cm} p{1cm} p{0.9cm} p{1.5cm}} 
    \\
    \toprule
    Dataset   & Length & Targeted Residues & Design Space & MSA Size & Train Fitness & Test Fitness & Reference \\
    \midrule
   \textbf{TrpB} \newline Enzyme \newline Activity  & 389 & 117, 118, 119, 162, 166, 182, 183, 184, 185, 186, 227, 228, 230, 231, 301 & \textit{N}=15 & $5.7e4$ & 75,618 & 23,313 & \citet{johnston_combinatorially_2024} \\
   \midrule
    \textbf{CreiLOV} Fluorescence & 119 & All & \textit{N}=119 & $3.7e5$ & 6,842 & 2,401 &\citet{chen_deep_2023} \\
    \midrule
    \textbf{GB1} Binding & 56 & All & \textit{N}=56 & $126$ & 3.9e6 & 9.6e4 &\citet{olson_comprehensive_2014} \\
    \bottomrule
  \end{tabular}
  \label{table:data}
\end{table}

We study three proteins, the TrpB enzyme \citep{johnston_combinatorially_2024}, the CreiLOV fluorescent protein \citep{chen_deep_2023}, and the GB1 binding protein \citep{olson_comprehensive_2014} due to the availability of fitness data across many residues (Table \ref{table:data}). We focus protein fitness optimization to a design space of 15 residues in TrpB (only these positions are allowed to vary) and all 119 and 56 residues in CreiLOV and GB1, respectively. For each protein's variants, we evaluate fitness by approximating it via a supervised oracle trained on a large amount of real data (Section \ref{section:task}).

\begin{table}[h]
  \caption{\textbf{Summary of generative priors evaluated in this work.} Each generative prior was trained on an MSA of homologous natural sequences. All denoising processes were modeled using a transformer architecture (Section \ref{section:model_methods}). \textit{\textbf{Italicized}} models were further explored in downstream guidance experiments.}
  \vspace{-0.2in}
  \centering
  \begin{tabular}
  {p{1.4cm}  p{1.4cm} p{1.4cm}  p{1.3cm} p{5.3cm}} 
    \\
    \toprule
    Model   & Type & Noise & \# Params  & Notes \\
    \midrule
    \textit{\textbf{Continuous}} & \multirow[c]{2}{*}{\shortstack[l]{Continuous\\Diffusion}} & \multirow[t]{2}{*}{\shortstack[l]{Gaussian}} & 27.9 M \\
    Continuous-ESM &  &  & 25.5 M & diffusion over ESM embeddings \\
    \midrule
     D3PM-Baseline & \multirow[c]{5}{*}{\shortstack[l]{Discrete\\Diffusion}} & \multirow[c]{5}{*}{\shortstack[l]{Uniform}} & 37.9 M  \\
   \textit{\textbf{D3PM}} &  &  & 37.9 M & finetuned from EvoDiff 38M-Uniform \citep{alamdari_protein_2023}\\
   UDLM &  & & 28.6 M & uniform diffusion langauge model \\
   \midrule
   \textit{\textbf{MDLM}} & Discrete Diffusion & Absorbing & 28.6 M & masked diffusion language model \\
    \midrule
   \textit{\textbf{ARLM}} & Language Model & n/a & 151 M  & autoregressive language model finetuned from ProGen2-small \citep{nijkamp_progen2_2023}\\
    \bottomrule
  \end{tabular}
  \vspace{-0in}
  \label{table:prior_training}
\end{table}

\subsection{Model pretraining captures the distribution of evolutionarily related protein sequences and enables sampling sequences with high fitness} 
\begin{figure}[h]
  \centering
\includegraphics[width=13cm]{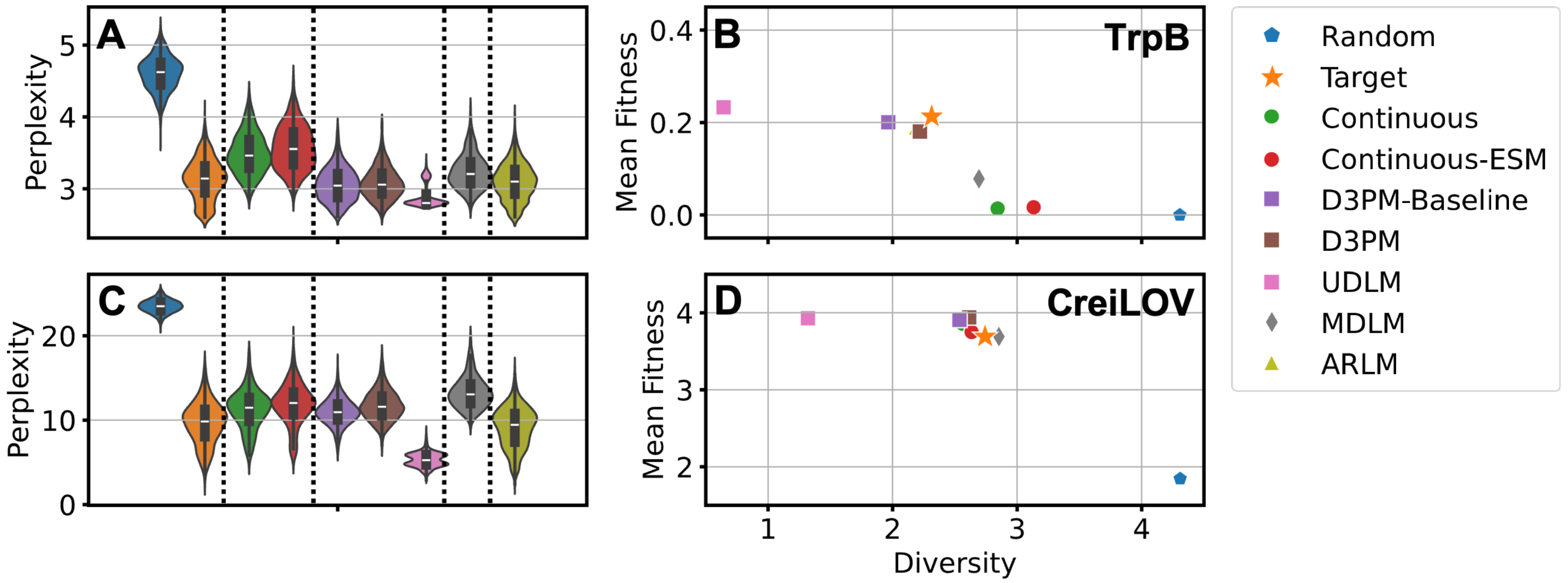}
  \vspace{-0.15in}
  \caption{\textbf{Pretrained generative priors capture the target distribution of naturally occurring sequences} that are homologous to TrpB (\textbf{A-B}) and CreiLOV (\textbf{C-D}), respectively. Lower perplexity corresponds to higher likelihood in the model. The diversity of sequences was computed as the average Shannon entropy of mutated positions with mean fitness corresponding to the oracle predictions. While the various models largely achieve comparable performance, the D3PM models capture the target distribution with the highest fidelity, whereas the UDLM model is prone to mode collapse. For each model, 1000 sequences were sampled and repeats were allowed to approximate the distribution. To approximate the target distribution, 1000 sequences were sampled from the MSA used for pretraining. Perplexity was calculated by passing generated sequences through the 764 M parameter ProGen2-base model. More details on model training can be found in Table \ref{table:prior_training} and Section \ref{section:model_methods}, and GB1 results are provided in Fig. \ref{fig:GB1_perplexity_logoplot}.} 
    \vspace{-0.2in}
  \label{fig:perplexity}
\end{figure}

Based on the methods explained in Section \ref{section:pretraining_data} and \ref{section:model_methods}, we trained generative priors on natural sequences from the MSA, focusing on continuous diffusion models (\textit{Continuous)}, discrete diffusion models with uniform (\textit{D3PM, UDLM}) and absorbing state noising processes (\textit{MDLM}), and autoregressive language models (\textit{ARLM}) (Table \ref{table:prior_training}). Overall, the trained models capture the natural distribution of protein sequences, with the D3PM models seeming to match the distribution the most closely while also generating sequences with high diversity (Fig. \ref{fig:perplexity}, Fig. \ref{fig:logoplot}). The two different diffusion models over continuous space show comparatively lower performance, and diffusing over the latent space of ESM embeddings does not boost performance on this task. The UDLM model has low performance due to mode collapse (Fig. \ref{fig:logoplot}, Fig. \ref{fig:hamming_distance}). Future work could finetune the pretrained diffusion protein language model (DPLM) as an MDLM \citep{wang_diffusion_2024}.

Overall, we found that pretrained priors sample protein variants that have higher mean fitness, which corroborates previous studies finding that sequences with higher evolutionary likelihood are also likely to have higher fitness \citep{li_evaluation_2024, hie_efficient_2023}. Based on these results, we proceeded to perform remaining experiments with one model from each category of model type, namely the \textit{Continuous}, \textit{D3PM}, \textit{MDLM}, and \textit{ARLM} models.

\begin{figure}[h]
  \centering
\includegraphics[width=\textwidth]{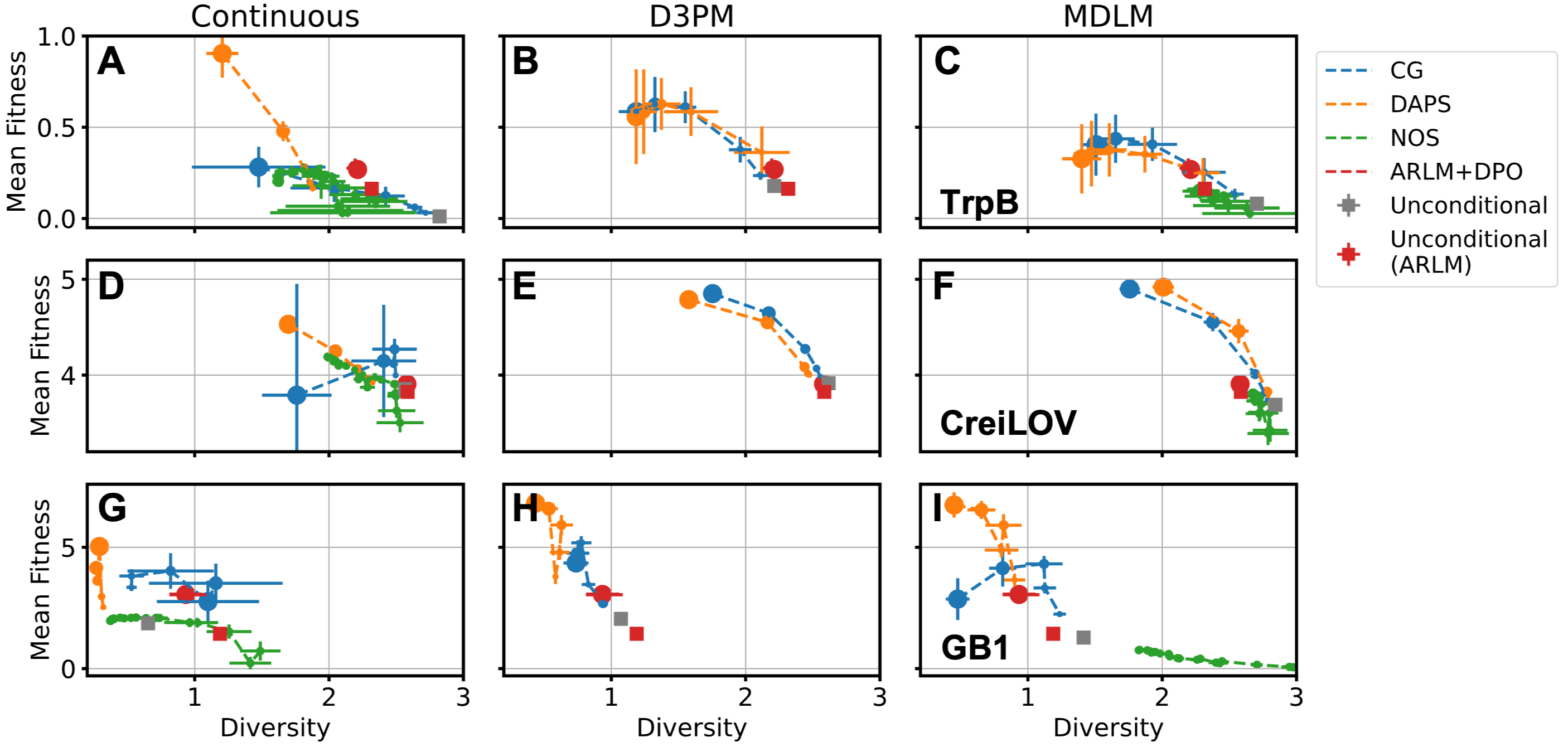}
  \vspace{-0.3in}
  \caption{\textbf{Pareto boundaries demonstrate the trade-off between generating sequences with high fitness and high diversity} for TrpB (\textbf{A-C}), CreiLOV (\textbf{D-F}), and GB1 (\textbf{G-I}). Sequences sampled from the generative models (Continuous, D3PM, and MDLM), after guidance with labeled fitness data, are enriched in high-fitness protein variants, and most methods show higher performance than the ARLM+DPO baseline. Larger circle indicates a stronger guidance strength hyperparameter (excluding NOS), specified in Table \ref{table:guidance_parameters}. Each experiment was repeated using 10 different standardized sets of 200 unique sequences used for steering, each drawn from the D3PM prior, and error bars show standard deviation. Mean fitness and diversity were calculated based on 200 generated samples, with diversity calculated as the average Shannon entropy of amino acids at mutated positions. Unconditional refers to sequences sampled from the prior with no guidance.}
  \vspace{-0.1in}
  \label{fig:fitness}
\end{figure}

\subsection{Evaluating SGPO design choices} 

Impressively, steering with modest amounts of labeled data (200 sequence-fitness pairs) enables most models and methods to generate sequences with even higher fitness, while sacrificing some generation diversity (Fig. \ref{fig:fitness}). In this low data regime, guidance with diffusion models outperforms DPO with language models; the latter does not enable as much steerability. CG and DAPS enable the strongest steerability overall, but DAPS outperforms CG for the continuous models (Fig. \ref{fig:fitness}A, D). In general, guidance seems to work similarly for uniform diffusion (D3PM) and to absorbing state diffusion (MDLM). Overall, the continuous diffusion models do not perform as well as other models, as the prior does not capture the distribution of natural sequences with high fitness as well (Fig. \ref{fig:fitness}A). NOS does not seem to allow for as much steerability, despite an extensive hyperparameter scan (Table \ref{table:guidance_parameters}). Finally, we conducted a closer analysis of the number of unique sequences generated by the steered models and confirmed that most models produce entirely novel sequences, suggesting that they are not over-steering (Fig. \ref{fig:pareto_unique}).

\subsection{``Thompson sampling'' using an ensemble of classifiers is effective for adaptive optimization}
\label{section:iterative}

Next, we performed adaptive optimization experiments, which mimic real-world protein engineering scenarios and follow a setup similar to batch Bayesian optimization: in each round, a batch of sequences is sampled, evaluated for fitness, and used to retrain a supervised value function that guides sampling from the pretrained prior. We focused on the MDLM models with the CG and DAPS guidance strategies, as these combinations achieved the best performance  in our earlier set of experiments (Fig. \ref{fig:fitness}). Based on findings from these previous experiments, we selected the ideal guidance strength hyperparameter to balance fitness and diversity—ensuring high predicted fitness without significantly compromising sequence diversity (Table \ref{table:guidance_parameters}). For both guidance strategies, we employed an algorithm akin to Thompson sampling \citep{kandasamy2018parallelised,russo2018tutorial}, drawing a different value function from a frequentist ensemble of neural network regressors to guide the generation of each new sample \citep{yang_active_2025}. Pseudocode for our adaptive optimization algorithm is provided in Section~\ref{section:adaptive_optimization}.

\begin{figure}[h]
  \centering
\includegraphics[width=\textwidth]{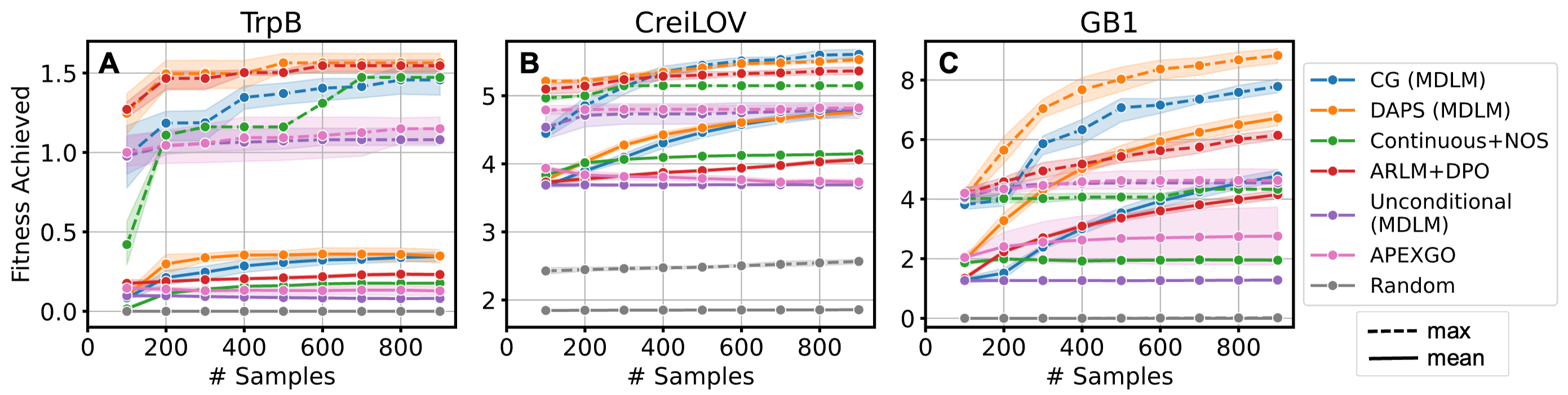}
  \vspace{-0.2in}
  \caption{\textbf{Maximum/mean fitness achieved improves over multiple iterations of steering in an adaptive setting similar to batch Bayesian optimization} for TrpB (\textbf{A}), CreiLOV (\textbf{B}), and GB1 (\textbf{C}). 100 sequences were sampled in each round. Within each round, an ensemble of 10 value functions (classifiers) was trained on fitness data from all previously queried samples, and each new sample was generated by the MDLM model guided with a value function sampled from the ensemble (akin to Thompson sampling). Only unique, novel samples were acquired. Guidance strength parameter is provided in Table \ref{table:guidance_parameters}. Error bars show standard deviation between 5 different random initializations.}
  \vspace{-0 in}
  \label{fig:iterative}
\end{figure}

Plug-and-play guidance strategies outperform baselines such as DPO with an ARLM, sampling just from the unconditional generative prior, and latent space Bayesian optimization with APEXGo  (Fig. \ref{fig:iterative}): Sampled sequences achieve higher values of mean and maximum fitness. Furthermore, campaigns using an ensemble of value functions and ``Thompson sampling'' achieve higher maximum fitness than those using only a single value function for guidance (Table \ref{table:ensemble_vs_single}), which may be because these models enable more exploration of sequence space (Fig. \ref{fig:diversity_ensemble}). However, it is difficult to ascertain wither CG or DAPS works better as a guidance strategy, as the performance is highly dependent on the guidance strength hyperparameter, and the optimal hyperparameter will not typically be known in a real-world campaign. Because the oracle may not capture the true nature of the protein fitness landscape, we also suggest making relative comparisons here rather than absolute comparisons between model performance.  

\section{Discussion}

In this work, we conduct a comprehensive study of SGPO methods and demonstrate that it is an effective approach for protein fitness optimization, by capturing the distribution of natural protein sequences with a generative prior and then steering the generations with labeled data. We find that DAPS with discrete diffusion models has the highest performance overall, and plug-and-play guidance-based strategies are generally more effective than finetuning language models; the latter can be difficult when only few fitness labels are available. SGPO approaches also outperform latent space Bayesian optimization (namely APEXGo), which we attribute to the difficulty in calibrating the trust region in very low-data regimes with limited rounds of optimization and the fact that latent space Bayesian optimization relies heavily on the structure of the latent space learned during generation model training, which can limit extrapolation to high-fitness but unnatural variants. 

Using plug-and-play guidance approaches has other advantages. First, only one hyperparameter (guidance strength) needs to be tuned. In real-world engineering scenarios, even in the absence of ground truth fitness labels, one practical approach to selecting the guidance strength is to scan over values and choose the highest setting for which $n$ generated sequences remain unique and novel relative to previously measured sequences, where $n$ corresponds to the screening throughput available for the next round. 
By contrast, for DPO, various hyperparameters need to be tuned, and the training process has to be monitored closely. Even for NOS, different parameters such as the step size, the number of steps, and the stability coefficient must be tuned together. A further advantage of guidance is the low computational cost required, as the prior model weights are not updated during guidance. Pretraining/finetuning to obtain each initial prior was achieved on a single H100 GPU in less than one hour while each individual guidance experiment took minutes; pretraining language models took several hours on a single GPU.   

Still, there are certain limitations of our work. We focused on proteins with fitness as mostly native function, but it would be interesting to test SGPO on other protein fitness optimization tasks where the pretrained prior may not provide as much utility. We also focused on protein optimization where only $\approx10^2$ fitness labels were available; different methods, such as RL, may perform better for applications where larger amounts of fitness data are available \citep{hie_adaptive_2022, blalock_functional_nodate}.  We focused on guidance strategies and did not test DPO or model finetuning-based methods with discrete diffusion models, but future work could adapt these methods for discrete diffusion \citep{borso_preference-based_2025}. Furthermore, for TrpB and for language models, we manually mapped sequences back into the design space after generation (Section \ref{section:task}), but explicitly building this into sampling techniques, such as inpainting in masked models \citep{blalock_functional_nodate, goel_memdlm_2024} may lead to improved performance. We did not consider insertions or deletions, but variable-length sequence generation could be considered in the future. Finally, we did not directly compare to existing approaches for protein engineering such as directed evolution for reasons explained in Section \ref{section:task}. 

There are several promising directions for future work to improve and extend SGPO methods. For instance, we experimented with guiding generation using value functions sampled from a Gaussian process posterior, enabling principled Thompson sampling from a fully Bayesian perspective. However, the Gaussian process struggled to model high-dimensional protein representations, leading to poor performance.  This limitation could potentially be addressed with better kernel choices \citep{wilson_deep_2015, michael_systematic_2024, yang_active_2025}. Recent work has also begun to incorporate multi-objective optimization \citep{annadani_preference-guided_2025, tang_peptune_2024, li_diffusion_2024, chen2025areurediannealedrectifiedupdates} and uncertainty quantification \citep{wu_diffusion-bbo_2025} when guiding diffusion models. Simultaneously, alternatives to acquisition-function-based approaches are being developed to enable Bayesian optimization in large design spaces where enumeration is infeasible \citep{bal_optimistic_2024}. Other emerging approaches—closer in spirit to flow matching—are being proposed for discrete data and may offer new opportunities for exploration \citep{davis_fisher_2024, stark_dirichlet_2024, tang_gumbel-softmax_2025}. Finally, for masked diffusion models, strategies such as remasking or scheduling could be explored to improve inference, particularly to enhance model amenability to guidance \citep{wang_remasking_2025, peng_path_2025, liu_think_2024, amin_why_2025}. It will also be interesting to further explore guidance in other discrete domains such as natural language and small molecules \citep{schiff_simple_2024}.

In summary, guiding generative models with labeled data offers a powerful, flexible, and principled framework for protein fitness optimization, as it effectively leverages both the evolutionary information encoded in natural protein sequences and task-specific fitness objectives. At the same time, we recognize the potential dual-use risks: such methods could, in principle, be misused to design harmful proteins, underscoring the importance of appropriate safeguards \citep{baker_protein_2024, wittmann_toward_2024}. In short, our work has examined multiple effective SGPO strategies and offered insights on best-practices for real-world protein fitness optimization, laying the groundwork for further exploration and wet-lab validation. 

\subsection*{Acknowledgments}
This work was supported by a U.S. Army Research Office cooperative agreement (W911NF-19-2-0026 to F.H.A.)  and an Amgen Chem-Bio Engineering award. J.Y. is also supported by the NSF Graduate Research Fellowship Program and the Google PhD Fellowship. We would like to thank Hunter Nisonoff, Jacob Gershon, Lucas Arnoldt, and Chenghao Liu for helpful discussions and Francesca-Zhoufan Li for help with the TrpB dataset. We would also like to thank Nate Gruver for help with the NOS implementation, Filippo Stocco for help with the DPO implementation, and Nathaniel Blalock for guidance on how to use the CreiLOV dataset. 

\bibliography{main}

\begin{thebibliography}{130}
\providecommand{\natexlab}[1]{#1}
\providecommand{\url}[1]{\texttt{#1}}
\expandafter\ifx\csname urlstyle\endcsname\relax
  \providecommand{\doi}[1]{doi: #1}\else
  \providecommand{\doi}{doi: \begingroup \urlstyle{rm}\Url}\fi

\bibitem[Alamdari et~al.(2024)Alamdari, Thakkar, van~den Berg, Tenenholtz, Strome, Moses, Lu, Fusi, Amini, and Yang]{alamdari_protein_2023}
Sarah Alamdari, Nitya Thakkar, Rianne van~den Berg, Neil Tenenholtz, Bob Strome, Alan~M. Moses, Alex~Xijie Lu, Nicol{\`o} Fusi, Ava~Pardis Amini, and Kevin~K. Yang.
\newblock Protein generation with evolutionary diffusion: sequence is all you need.
\newblock \emph{bioRxiv}, November 2024.
\newblock \doi{10.1101/2023.09.11.556673}.
\newblock URL \url{https://www.biorxiv.org/content/10.1101/2023.09.11.556673v2}.
\newblock Preprint; version 2 posted Nov 4, 2024.

\bibitem[Amin et~al.(2025{\natexlab{a}})Amin, Gruver, and Wilson]{amin_why_2025}
Alan~N. Amin, Nate Gruver, and Andrew~Gordon Wilson.
\newblock Why masking diffusion works: Condition on the jump schedule for improved discrete diffusion.
\newblock In \emph{Advances in Neural Information Processing Systems (NeurIPS)}, 2025{\natexlab{a}}.
\newblock \doi{10.48550/arXiv.2506.08316}.
\newblock URL \url{https://neurips.cc/virtual/2025/poster/115376}.
\newblock NeurIPS 2025 (poster), to appear.

\bibitem[Amin et~al.(2025{\natexlab{b}})Amin, Gruver, Kuang, Li, Elliott, McCarter, Raghu, Greenside, and Wilson]{amin_bayesian_2024}
Alan~Nawzad Amin, Nate Gruver, Yilun Kuang, Yucen~Lily Li, Hunter Elliott, Calvin McCarter, Aniruddh Raghu, Peyton Greenside, and Andrew~Gordon Wilson.
\newblock Bayesian optimization of antibodies informed by a generative model of evolving sequences.
\newblock In \emph{Proceedings of the Thirteenth International Conference on Learning Representations (ICLR)}, Singapore, April 2025{\natexlab{b}}.
\newblock \doi{10.48550/arXiv.2412.07763}.
\newblock URL \url{https://openreview.net/forum?id=E48QvQppIN}.
\newblock Spotlight.

\bibitem[Angermueller et~al.(2020)Angermueller, Dohan, Belanger, Deshpande, Murphy, and Colwell]{angermueller_model-based_2020}
Christof Angermueller, David Dohan, David Belanger, Ramya Deshpande, Kevin Murphy, and Lucy Colwell.
\newblock Model-based reinforcement learning for biological sequence design.
\newblock In \emph{Proceedings of the 8th International Conference on Learning Representations (ICLR)}, 2020.
\newblock URL \url{https://openreview.net/forum?id=HklxbgBKvr}.
\newblock ICLR 2020.

\bibitem[Annadani et~al.(2025)Annadani, Belakaria, Ermon, Bauer, and Engelhardt]{annadani_preference-guided_2025}
Yashas Annadani, Syrine Belakaria, Stefano Ermon, Stefan Bauer, and Barbara~E. Engelhardt.
\newblock Preference-guided diffusion for multi-objective offline optimization.
\newblock In \emph{Advances in Neural Information Processing Systems (NeurIPS)}, 2025.
\newblock \doi{10.48550/arXiv.2503.17299}.
\newblock URL \url{https://neurips.cc/virtual/2025/poster/116185}.
\newblock Poster; to appear.

\bibitem[Arnold(2018)]{arnold_directed_2018}
Frances~H. Arnold.
\newblock Directed evolution: Bringing new chemistry to life.
\newblock \emph{Angewandte Chemie International Edition}, 57\penalty0 (16):\penalty0 4143--4148, April 2018.
\newblock \doi{10.1002/anie.201708408}.
\newblock URL \url{https://onlinelibrary.wiley.com/doi/10.1002/anie.201708408}.

\bibitem[Austin et~al.(2021)Austin, Johnson, Ho, Tarlow, and van~den Berg]{austin_structured_2023}
Jacob Austin, Daniel~D. Johnson, Jonathan Ho, Daniel Tarlow, and Rianne van~den Berg.
\newblock Structured denoising diffusion models in discrete state-spaces.
\newblock In \emph{Advances in Neural Information Processing Systems}, volume~34, 2021.
\newblock URL \url{https://proceedings.neurips.cc/paper/2021/hash/958c530554f78bcd8e97125b70e6973d-Abstract.html}.
\newblock NeurIPS 2021.

\bibitem[Baker \& Church(2024)Baker and Church]{baker_protein_2024}
David Baker and George Church.
\newblock Protein design meets biosecurity.
\newblock \emph{Science}, 383\penalty0 (6681):\penalty0 349, January 2024.
\newblock \doi{10.1126/science.ado1671}.
\newblock URL \url{https://www.science.org/doi/10.1126/science.ado1671}.

\bibitem[Bal et~al.(2025)Bal, Sessa, Mutny, and Krause]{bal_optimistic_2024}
Melis~Ilayda Bal, Pier~Giuseppe Sessa, Mojmir Mutny, and Andreas Krause.
\newblock Optimistic games for combinatorial bayesian optimization with application to protein design.
\newblock In \emph{Proceedings of the Thirteenth International Conference on Learning Representations (ICLR)}, Singapore, April 2025.
\newblock \doi{10.48550/arXiv.2409.18582}.
\newblock URL \url{https://openreview.net/forum?id=xiyzCfXTS6}.
\newblock Poster.

\bibitem[Bartoldson et~al.(2025)Bartoldson, Venkatraman, Diffenderfer, Jain, Ben-Nun, Lee, Kim, Obando-Ceron, Bengio, and Kailkhura]{bartoldson_trajectory_2025}
Brian~R. Bartoldson, Siddarth Venkatraman, James Diffenderfer, Moksh Jain, Tal Ben-Nun, Seanie Lee, Minsu Kim, Johan Obando-Ceron, Yoshua Bengio, and Bhavya Kailkhura.
\newblock Trajectory balance with asynchrony: Decoupling exploration and learning for fast, scalable llm post-training.
\newblock In \emph{Advances in Neural Information Processing Systems (NeurIPS)}, 2025.
\newblock \doi{10.48550/arXiv.2503.18929}.
\newblock URL \url{https://neurips.cc/virtual/2025/poster/117641}.
\newblock Poster; to appear.

\bibitem[Blalock et~al.(2025)Blalock, Seshadri, Babbar, Fahlberg, Kulkarni, and Romero]{blalock_functional_nodate}
Nathaniel Blalock, Srinath Seshadri, Agrim Babbar, Sarah~A. Fahlberg, Ameya Kulkarni, and Philip~A. Romero.
\newblock Functional alignment of protein language models via reinforcement learning.
\newblock \emph{bioRxiv}, May 2025.
\newblock \doi{10.1101/2025.05.02.651993}.
\newblock URL \url{https://www.biorxiv.org/content/10.1101/2025.05.02.651993v1}.
\newblock Preprint.

\bibitem[Borso et~al.(2025)Borso, Paglieri, Wells, and Rockt{\"a}schel]{borso_preference-based_2025}
Umberto Borso, Davide Paglieri, Jude Wells, and Tim Rockt{\"a}schel.
\newblock Preference-based alignment of discrete diffusion models.
\newblock In \emph{ICLR 2025 Workshop on Bidirectional Human--AI Alignment (Bi-Align)}, Singapore, April 2025.
\newblock \doi{10.48550/arXiv.2503.08295}.
\newblock URL \url{https://openreview.net/pdf?id=qs9CTsC32h}.
\newblock Workshop paper.

\bibitem[Brookes \& Listgarten(2020)Brookes and Listgarten]{brookes_design_2020}
David~H. Brookes and Jennifer Listgarten.
\newblock Design by adaptive sampling, February 2020.
\newblock URL \url{http://arxiv.org/abs/1810.03714}.
\newblock arXiv:1810.03714 [cs, q-bio, stat].

\bibitem[Brookes et~al.(2019)Brookes, Park, and Listgarten]{brookes_conditioning_nodate}
David~H Brookes, Hahnbeom Park, and Jennifer Listgarten.
\newblock Conditioning by adaptive sampling for robust design.
\newblock In \emph{International Conference on Machine Learning}, 2019.
\newblock URL \url{https://arxiv.org/abs/1901.10060}.

\bibitem[Campbell et~al.(2022)Campbell, Benton, De~Bortoli, Rainforth, Deligiannidis, and Doucet]{campbell_continuous_2022}
Andrew Campbell, Joe Benton, Valentin De~Bortoli, Tom Rainforth, George Deligiannidis, and Arnaud Doucet.
\newblock A continuous time framework for discrete denoising models.
\newblock In \emph{Advances in Neural Information Processing Systems (NeurIPS)}, volume~35, pp.\  28266--28279, New Orleans, LA, USA, December 2022. Curran Associates, Inc.
\newblock URL \url{https://proceedings.neurips.cc/paper_files/paper/2022/hash/b5b528767aa35f5b1a60fe0aaeca0563-Abstract-Conference.html}.
\newblock NeurIPS 2022.

\bibitem[Campbell et~al.(2024)Campbell, Yim, Barzilay, Rainforth, and Jaakkola]{campbell_generative_2024}
Andrew Campbell, Jason Yim, Regina Barzilay, Tom Rainforth, and Tommi Jaakkola.
\newblock Generative flows on discrete state-spaces: Enabling multimodal flows with applications to protein co-design.
\newblock In \emph{Proceedings of the 41st International Conference on Machine Learning (ICML)}, volume 235 of \emph{Proceedings of Machine Learning Research}, pp.\  5453--5512. PMLR, 21--27 Jul 2024.
\newblock URL \url{https://proceedings.mlr.press/v235/campbell24a.html}.

\bibitem[Castro et~al.(2022)Castro, Godavarthi, Rubinfien, Givechian, Bhaskar, and Krishnaswamy]{castro_guided_2022}
Egbert Castro, Abhinav Godavarthi, Julian Rubinfien, Kevin~B. Givechian, Dhananjay Bhaskar, and Smita Krishnaswamy.
\newblock Transformer-based protein generation with regularized latent space optimization.
\newblock \emph{Nature Machine Intelligence}, 4\penalty0 (10):\penalty0 840--851, October 2022.
\newblock \doi{10.1038/s42256-022-00532-1}.
\newblock URL \url{https://www.nature.com/articles/s42256-022-00532-1}.

\bibitem[Chen et~al.(2023{\natexlab{a}})Chen, Vure, Pulugurta, and Chatterjee]{chen_amp-diffusion_2024}
Tianlai Chen, Pranay Vure, Rishab Pulugurta, and Pranam Chatterjee.
\newblock Amp-diffusion: Integrating latent diffusion with protein language models for antimicrobial peptide generation.
\newblock In \emph{NeurIPS 2023 Workshop on Generative AI and Biology (GenBio)}, New Orleans, LA, USA, December 2023{\natexlab{a}}.
\newblock URL \url{https://openreview.net/forum?id=145TM9VQhx}.
\newblock Workshop poster; non-archival.

\bibitem[Chen et~al.(2023{\natexlab{b}})Chen, Zhang, and Hinton]{chen_analog_2023}
Ting Chen, Ruixiang Zhang, and Geoffrey~E. Hinton.
\newblock Analog bits: Generating discrete data using diffusion models with self-conditioning.
\newblock In \emph{Proceedings of the Eleventh International Conference on Learning Representations (ICLR)}, May 2023{\natexlab{b}}.
\newblock URL \url{https://openreview.net/forum?id=3itjR9QxFw}.
\newblock ICLR 2023.

\bibitem[Chen et~al.(2025)Chen, Zhang, and Chatterjee]{chen2025areurediannealedrectifiedupdates}
Tong Chen, Yinuo Zhang, and Pranam Chatterjee.
\newblock Areuredi: Annealed rectified updates for refining discrete flows with multi-objective guidance, 2025.
\newblock URL \url{https://arxiv.org/abs/2510.00352}.

\bibitem[Chen et~al.(2023{\natexlab{c}})Chen, Hu, Li, Zhang, Fu, Zhang, and Si]{chen_deep_2023}
Yongcan Chen, Ruyun Hu, Keyi Li, Yating Zhang, Lihao Fu, Jianzhi Zhang, and Tong Si.
\newblock Deep mutational scanning of an oxygen-independent fluorescent protein creilov for comprehensive profiling of mutational and epistatic effects.
\newblock \emph{ACS Synthetic Biology}, 12\penalty0 (5):\penalty0 1461--1473, May 2023{\natexlab{c}}.
\newblock \doi{10.1021/acssynbio.2c00662}.
\newblock URL \url{https://pubs.acs.org/doi/10.1021/acssynbio.2c00662}.

\bibitem[Chu et~al.(2025)Chu, Wu, Chen, Song, and Yue]{chu2025split}
Wenda Chu, Zihui Wu, Yifan Chen, Yang Song, and Yisong Yue.
\newblock Split gibbs discrete diffusion posterior sampling.
\newblock In \emph{Advances in Neural Information Processing Systems (NeurIPS)}, 2025.
\newblock \doi{10.48550/arXiv.2503.01161}.
\newblock URL \url{https://neurips.cc/virtual/2025/poster/117795}.
\newblock Poster; to appear.

\bibitem[Chung et~al.(2023)Chung, Kim, Mccann, Klasky, and Ye]{chung2023diffusion}
Hyungjin Chung, Jeongsol Kim, Michael~Thompson Mccann, Marc~Louis Klasky, and Jong~Chul Ye.
\newblock Diffusion posterior sampling for general noisy inverse problems.
\newblock In \emph{The Eleventh International Conference on Learning Representations}, 2023.
\newblock URL \url{https://openreview.net/forum?id=OnD9zGAGT0k}.

\bibitem[Dathathri et~al.(2020)Dathathri, Madotto, Lan, Hung, Frank, Molino, Yosinski, and Liu]{dathathri_plug_2020}
Sumanth Dathathri, Andrea Madotto, Janice Lan, Jane Hung, Eric Frank, Piero Molino, Jason Yosinski, and Rosanne Liu.
\newblock Plug and play language models: A simple approach to controlled text generation.
\newblock In \emph{Proceedings of the 8th International Conference on Learning Representations (ICLR)}, 2020.
\newblock URL \url{https://openreview.net/forum?id=H1edEyBKDS}.
\newblock ICLR 2020.

\bibitem[Davis et~al.(2024)Davis, Kessler, Petrache, Ceylan, Bronstein, and Bose]{davis_fisher_2024}
Oscar Davis, Samuel Kessler, Mircea Petrache, Ismail~Ilkan Ceylan, Michael Bronstein, and Avishek~Joey Bose.
\newblock Fisher flow matching for generative modeling over discrete data.
\newblock In \emph{Advances in Neural Information Processing Systems (NeurIPS)}, volume~37, 2024.
\newblock URL \url{https://proceedings.neurips.cc/paper_files/paper/2024/hash/fadec8f2e65f181d777507d1df69b92f-Abstract-Conference.html}.

\bibitem[Dieleman et~al.(2022)Dieleman, Sartran, Roshannai, Savinov, Ganin, Richemond, Doucet, Strudel, Dyer, Durkan, Hawthorne, Leblond, Grathwohl, and Adler]{dieleman_continuous_2022}
Sander Dieleman, Laurent Sartran, Arman Roshannai, Nikolay Savinov, Yaroslav Ganin, Pierre~H. Richemond, Arnaud Doucet, Robin Strudel, Chris Dyer, Conor Durkan, Curtis Hawthorne, R{\'e}mi Leblond, Will Grathwohl, and Jonas Adler.
\newblock Continuous diffusion for categorical data.
\newblock \emph{arXiv preprint arXiv:2211.15089}, November 2022.
\newblock \doi{10.48550/arXiv.2211.15089}.
\newblock URL \url{https://arxiv.org/abs/2211.15089}.
\newblock Preprint.

\bibitem[Ding et~al.(2024)Ding, Chin, Zhao, Huang, Mai, Wang, Liu, Yang, and Luo]{ding_machine_2024}
Kerr Ding, Michael Chin, Yunlong Zhao, Wei Huang, Binh~Khanh Mai, Huanan Wang, Peng Liu, Yang Yang, and Yunan Luo.
\newblock Machine learning-guided co-optimization of fitness and diversity facilitates combinatorial library design in enzyme engineering.
\newblock \emph{Nature Communications}, 15\penalty0 (1):\penalty0 6392, July 2024.
\newblock \doi{10.1038/s41467-024-50698-y}.
\newblock URL \url{https://www.nature.com/articles/s41467-024-50698-y}.

\bibitem[Fei et~al.(2025)Fei, Li, Liu, Wei, Chen, and Gao]{fei_advancing_2025}
Hongyuan Fei, Yunjia Li, Yijing Liu, Jingjing Wei, Aojie Chen, and Caixia Gao.
\newblock Advancing protein evolution with inverse folding models integrating structural and evolutionary constraints.
\newblock \emph{Cell}, 188\penalty0 (17):\penalty0 4674--4692.e19, August 2025.
\newblock \doi{10.1016/j.cell.2025.06.014}.
\newblock URL \url{https://www.cell.com/cell/abstract/S0092-8674(25)00680-4}.

\bibitem[Frazier(2018)]{frazier2018tutorial}
Peter~I. Frazier.
\newblock Bayesian optimization.
\newblock In \emph{Recent Advances in Optimization and Modeling of Contemporary Problems}, INFORMS TutORials in Operations Research, pp.\  255--278. INFORMS, Catonsville, MD, 2018.
\newblock \doi{10.1287/educ.2018.0188}.
\newblock URL \url{https://pubsonline.informs.org/doi/10.1287/educ.2018.0188}.

\bibitem[Freschlin et~al.(2022)Freschlin, Fahlberg, and Romero]{freschlin_machine_2022}
Chase~R Freschlin, Sarah~A Fahlberg, and Philip~A Romero.
\newblock Machine learning to navigate fitness landscapes for protein engineering.
\newblock \emph{Current Opinion in Biotechnology}, 75:\penalty0 102713, June 2022.
\newblock ISSN 0958-1669.
\newblock \doi{10.1016/j.copbio.2022.102713}.
\newblock URL \url{https://www.sciencedirect.com/science/article/pii/S0958166922000465}.

\bibitem[Goel et~al.(2025)Goel, Thoutam, Marroquin, Gokaslan, Firouzbakht, Vincoff, Kuleshov, Kratochvil, and Chatterjee]{goel_memdlm_2024}
Shrey Goel, Vishrut Thoutam, Edgar~Mariano Marroquin, Aaron Gokaslan, Arash Firouzbakht, Sophia Vincoff, Volodymyr Kuleshov, Huong~T. Kratochvil, and Pranam Chatterjee.
\newblock Memdlm: De novo membrane protein design with property-guided discrete diffusion.
\newblock In \emph{ICLR 2025 Workshop on Learning Meaningful Representations of Life (LMRL)}, Singapore, April 2025.
\newblock URL \url{https://iclr.cc/virtual/2025/35945}.
\newblock Workshop poster; non-archival.

\bibitem[G{\'o}mez-Bombarelli et~al.(2018)G{\'o}mez-Bombarelli, Wei, Duvenaud, Hern{\'a}ndez-Lobato, S{\'a}nchez-Lengeling, Sheberla, Aguilera-Iparraguirre, Hirzel, Adams, and Aspuru-Guzik]{gomez-bombarelli_automatic_2018}
Rafael G{\'o}mez-Bombarelli, Jennifer~N. Wei, David Duvenaud, Jos{\'e}~Miguel Hern{\'a}ndez-Lobato, Benjam{\'i}n S{\'a}nchez-Lengeling, Dennis Sheberla, Jorge Aguilera-Iparraguirre, Timothy~D. Hirzel, Ryan~P. Adams, and Al{\'a}n Aspuru-Guzik.
\newblock Automatic chemical design using a data-driven continuous representation of molecules.
\newblock \emph{ACS Central Science}, 4\penalty0 (2):\penalty0 268--276, February 2018.
\newblock \doi{10.1021/acscentsci.7b00572}.
\newblock URL \url{https://pubs.acs.org/doi/10.1021/acscentsci.7b00572}.

\bibitem[Gruver et~al.(2023)Gruver, Stanton, Frey, Rudner, Hotzel, Lafrance-Vanasse, Rajpal, Cho, and Wilson]{gruver_protein_2023}
Nate Gruver, Samuel Stanton, Nathan~C. Frey, Tim G.~J. Rudner, Isidro Hotzel, Julien Lafrance-Vanasse, Arvind Rajpal, Kyunghyun Cho, and Andrew~Gordon Wilson.
\newblock Protein design with guided discrete diffusion.
\newblock In \emph{Advances in Neural Information Processing Systems}, volume~36, 2023.
\newblock URL \url{https://proceedings.neurips.cc/paper_files/paper/2023/hash/29591f355702c3f4436991335784b503-Abstract-Conference.html}.
\newblock NeurIPS 2023.

\bibitem[Gupta \& Zou(2019)Gupta and Zou]{gupta_feedback_2019}
Anvita Gupta and James Zou.
\newblock Feedback {GAN} for {DNA} optimizes protein functions.
\newblock \emph{Nature Machine Intelligence}, 1\penalty0 (2):\penalty0 105--111, February 2019.
\newblock ISSN 2522-5839.
\newblock \doi{10.1038/s42256-019-0017-4}.
\newblock URL \url{https://www.nature.com/articles/s42256-019-0017-4}.
\newblock Number: 2 Publisher: Nature Publishing Group.

\bibitem[Hawkins-Hooker et~al.(2024)Hawkins-Hooker, Kmec, Bent, and Duckworth]{hawkins-hooker_likelihood-based_2024}
Alex Hawkins-Hooker, Jakub Kmec, Oliver Bent, and Paul Duckworth.
\newblock Likelihood-based fine-tuning of protein language models for few-shot fitness prediction and design.
\newblock In \emph{ICML 2024 Workshop on Machine Learning for Life and Material Science: From Theory to Industry Applications (ML4LMS)}, Vienna, Austria, July 2024.
\newblock \doi{10.1101/2024.05.28.596156}.
\newblock URL \url{https://openreview.net/forum?id=MkYhOEUJyi}.
\newblock Workshop poster; non-archival.

\bibitem[Hie \& Yang(2022)Hie and Yang]{hie_adaptive_2022}
Brian~L. Hie and Kevin~K. Yang.
\newblock Adaptive machine learning for protein engineering.
\newblock \emph{Current Opinion in Structural Biology}, 72:\penalty0 145--152, February 2022.
\newblock ISSN 0959440X.
\newblock \doi{10.1016/j.sbi.2021.11.002}.
\newblock URL \url{https://linkinghub.elsevier.com/retrieve/pii/S0959440X21001457}.

\bibitem[Hie et~al.(2023)Hie, Shanker, Xu, Bruun, Weidenbacher, Tang, Wu, Pak, and Kim]{hie_efficient_2023}
Brian~L. Hie, Varun~R. Shanker, Duo Xu, Theodora U.~J. Bruun, Payton~A. Weidenbacher, Shaogeng Tang, Wesley Wu, John~E. Pak, and Peter~S. Kim.
\newblock Efficient evolution of human antibodies from general protein language models.
\newblock \emph{Nature Biotechnology}, April 2023.
\newblock ISSN 1546-1696.
\newblock \doi{10.1038/s41587-023-01763-2}.
\newblock URL \url{https://doi.org/10.1038/s41587-023-01763-2}.

\bibitem[Ho et~al.(2020)Ho, Jain, and Abbeel]{ho_denoising_2020}
Jonathan Ho, Ajay Jain, and Pieter Abbeel.
\newblock Denoising diffusion probabilistic models.
\newblock In \emph{Advances in Neural Information Processing Systems (NeurIPS)}, volume~33, 2020.
\newblock URL \url{https://proceedings.neurips.cc/paper/2020/hash/4c5bcfec8584af0d967f1ab10179ca4b-Abstract.html}.
\newblock NeurIPS 2020.

\bibitem[Hoogeboom et~al.(2022)Hoogeboom, Gritsenko, Bastings, Poole, van~den Berg, and Salimans]{hoogeboom_autoregressive_2022}
Emiel Hoogeboom, Alexey~A. Gritsenko, Jasmijn Bastings, Ben Poole, Rianne van~den Berg, and Tim Salimans.
\newblock Autoregressive diffusion models.
\newblock In \emph{Proceedings of the Tenth International Conference on Learning Representations (ICLR)}, April 2022.
\newblock URL \url{https://openreview.net/forum?id=Lm8T39vLDTE}.
\newblock ICLR 2022 (poster).

\bibitem[Hsu et~al.(2022)Hsu, Nisonoff, Fannjiang, and Listgarten]{hsu_learning_2022}
Chloe Hsu, Hunter Nisonoff, Clara Fannjiang, and Jennifer Listgarten.
\newblock Learning protein fitness models from evolutionary and assay-labeled data.
\newblock \emph{Nature Biotechnology}, 40\penalty0 (7):\penalty0 1114--1122, January 2022.
\newblock ISSN 1546-1696.
\newblock \doi{10.1038/s41587-021-01146-5}.
\newblock URL \url{https://www.nature.com/articles/s41587-021-01146-5}.
\newblock Bandiera\_abtest: a Cg\_type: Nature Research Journals Primary\_atype: Research Publisher: Nature Publishing Group Subject\_term: Machine learning;Protein design Subject\_term\_id: machine-learning;protein-design.

\bibitem[Hsu et~al.(2024)Hsu, Fannjiang, and Listgarten]{hsu_generative_2024}
Chloe Hsu, Clara Fannjiang, and Jennifer Listgarten.
\newblock Generative models for protein structures and sequences.
\newblock \emph{Nature Biotechnology}, 42:\penalty0 196--199, 2024.

\bibitem[Huang et~al.(2025)Huang, Zhu, He, and Yao]{pmlr-v267-huang25ba}
Long-Kai Huang, Rongyi Zhu, Bing He, and Jianhua Yao.
\newblock Steering protein language models.
\newblock In Aarti Singh, Maryam Fazel, Daniel Hsu, Simon Lacoste-Julien, Felix Berkenkamp, Tegan Maharaj, Kiri Wagstaff, and Jerry Zhu (eds.), \emph{Proceedings of the 42nd International Conference on Machine Learning}, volume 267 of \emph{Proceedings of Machine Learning Research}, pp.\  26247--26260. PMLR, 13--19 Jul 2025.
\newblock URL \url{https://proceedings.mlr.press/v267/huang25ba.html}.

\bibitem[Jain et~al.(2022)Jain, Bengio, Hernandez-Garcia, Rector-Brooks, Dossou, Ekbote, Fu, Zhang, Kilgour, Zhang, Simine, Das, and Bengio]{jain_biological_2022}
Moksh Jain, Emmanuel Bengio, Alex Hernandez-Garcia, Jarrid Rector-Brooks, Bonaventure F.~P. Dossou, Chanakya Ekbote, Jie Fu, Tianyu Zhang, Michael Kilgour, Dinghuai Zhang, Lena Simine, Payel Das, and Yoshua Bengio.
\newblock Biological sequence design with {GFlowNets}.
\newblock In \emph{Proceedings of the 39th International Conference on Machine Learning (ICML)}, volume 162 of \emph{Proceedings of Machine Learning Research}, pp.\  9786--9801. PMLR, 17--23 Jul 2022.
\newblock URL \url{https://proceedings.mlr.press/v162/jain22a.html}.

\bibitem[Jiang et~al.(2024)Jiang, Yan, Di~Bernardo, Sgrizzi, Villiger, Kayabolen, Kim, Carscadden, Hiraizumi, Nishimasu, Gootenberg, and Abudayyeh]{jiang_rapid_2024}
Kaiyi Jiang, Zhaoqing Yan, Matteo Di~Bernardo, Samantha~R. Sgrizzi, Lukas Villiger, Alisan Kayabolen, B.~J. Kim, Josephine~K. Carscadden, Masahiro Hiraizumi, Hiroshi Nishimasu, Jonathan~S. Gootenberg, and Omar~O. Abudayyeh.
\newblock Rapid in silico directed evolution by a protein language model with {EVOLVEpro}.
\newblock \emph{Science}, 387\penalty0 (6732):\penalty0 eadr6006, November 2024.
\newblock \doi{10.1126/science.adr6006}.
\newblock URL \url{https://www.science.org/doi/10.1126/science.adr6006}.
\newblock Publisher: American Association for the Advancement of Science.

\bibitem[Johnson et~al.(2010)Johnson, Eddy, and Portugaly]{johnson_hidden_2010}
L.~Steven Johnson, Sean~R. Eddy, and Elon Portugaly.
\newblock Hidden {Markov} model speed heuristic and iterative {HMM} search procedure.
\newblock \emph{BMC Bioinformatics}, 11\penalty0 (1):\penalty0 431, August 2010.
\newblock ISSN 1471-2105.
\newblock \doi{10.1186/1471-2105-11-431}.
\newblock URL \url{https://doi.org/10.1186/1471-2105-11-431}.

\bibitem[Johnston et~al.(2024)Johnston, Almhjell, Watkins-Dulaney, Liu, Porter, Yang, and Arnold]{johnston_combinatorially_2024}
Kadina~E. Johnston, Patrick~J. Almhjell, Ella~J. Watkins-Dulaney, Grace Liu, Nicholas~J. Porter, Jason Yang, and Frances~H. Arnold.
\newblock A combinatorially complete epistatic fitness landscape in an enzyme active site.
\newblock \emph{Proceedings of the National Academy of Sciences}, 121\penalty0 (32):\penalty0 e2400439121, August 2024.
\newblock \doi{10.1073/pnas.2400439121}.
\newblock URL \url{https://www.pnas.org/doi/10.1073/pnas.2400439121}.

\bibitem[Kandasamy et~al.(2018)Kandasamy, Krishnamurthy, Schneider, and P{\'o}czos]{kandasamy2018parallelised}
Kirthevasan Kandasamy, Akshay Krishnamurthy, Jeff Schneider, and Barnab{\'a}s P{\'o}czos.
\newblock Parallelised bayesian optimisation via thompson sampling.
\newblock In \emph{International conference on artificial intelligence and statistics}, pp.\  133--142. PMLR, 2018.

\bibitem[Katoh \& Standley(2013)Katoh and Standley]{katoh_mafft_2013}
Kazutaka Katoh and Daron~M. Standley.
\newblock {MAFFT} {Multiple} {Sequence} {Alignment} {Software} {Version} 7: {Improvements} in {Performance} and {Usability}.
\newblock \emph{Molecular Biology and Evolution}, 30\penalty0 (4):\penalty0 772--780, April 2013.
\newblock ISSN 0737-4038.
\newblock \doi{10.1093/molbev/mst010}.
\newblock URL \url{https://doi.org/10.1093/molbev/mst010}.

\bibitem[Kim et~al.(2025)Kim, Kim, Yun, Choi, Bengio, Hern{\'a}ndez-Garc{\'\i}a, and Park]{kim_improved_2025}
Hyeonah Kim, Minsu Kim, Taeyoung Yun, Sanghyeok Choi, Emmanuel Bengio, Alex Hern{\'a}ndez-Garc{\'\i}a, and Jinkyoo Park.
\newblock Improved off-policy reinforcement learning in biological sequence design.
\newblock In \emph{Proceedings of the 42nd International Conference on Machine Learning (ICML)}, Vancouver, Canada, July 2025.
\newblock \doi{10.48550/arXiv.2410.04461}.
\newblock URL \url{https://icml.cc/virtual/2025/poster/46683}.
\newblock ICML 2025 (poster), to appear.

\bibitem[Kirjner et~al.(2024)Kirjner, Yim, Samusevich, Bracha, Jaakkola, Barzilay, and Fiete]{kirjner2024improvingproteinoptimizationsmoothed}
Andrew Kirjner, Jason Yim, Raman Samusevich, Shahar Bracha, Tommi Jaakkola, Regina Barzilay, and Ila Fiete.
\newblock Improving protein optimization with smoothed fitness landscapes.
\newblock In \emph{Proceedings of the 12th International Conference on Learning Representations}, 2024.
\newblock URL \url{https://arxiv.org/abs/2307.00494}.

\bibitem[Klarner et~al.(2024)Klarner, Rudner, Morris, Deane, and Teh]{klarner_context-guided_2024}
Leo Klarner, Tim G.~J. Rudner, Garrett~M. Morris, Charlotte~M. Deane, and Yee~Whye Teh.
\newblock Context-guided diffusion for out-of-distribution molecular and protein design.
\newblock In \emph{Proceedings of the 41st International Conference on Machine Learning (ICML)}, volume 235 of \emph{Proceedings of Machine Learning Research}, pp.\  24770--24807. PMLR, 21--27 Jul 2024.
\newblock URL \url{https://proceedings.mlr.press/v235/klarner24a.html}.

\bibitem[Lambert et~al.(2025)Lambert, Tavakoli, Dharuman, Yang, Bhethanabotla, Kaur, Hill, Ramanathan, Anandkumar, and Arnold]{lambert_sequence-based_nodate}
Th{\'e}ophile Lambert, Amin Tavakoli, Gautham Dharuman, Jason Yang, Vignesh Bhethanabotla, Sukhvinder Kaur, Matthew Hill, Arvind Ramanathan, Anima Anandkumar, and Frances~H. Arnold.
\newblock Sequence-based generative {AI}-guided design of versatile tryptophan synthases.
\newblock \emph{bioRxiv}, August 2025.
\newblock \doi{10.1101/2025.08.30.673177}.
\newblock URL \url{https://www.biorxiv.org/content/10.1101/2025.08.30.673177v1}.
\newblock Preprint; version 1 posted Aug 30, 2025.

\bibitem[Lee et~al.(2025{\natexlab{a}})Lee, Jeha, Frellsen, Li{\`o}, Albergo, and Vargas]{lee_debiasing_2025}
Cheuk~Kit Lee, Paul Jeha, Jes Frellsen, Pietro Li{\`o}, Michael~Samuel Albergo, and Francisco Vargas.
\newblock Debiasing guidance for discrete diffusion with sequential monte carlo.
\newblock In \emph{ICLR 2025 Workshop on Frontiers in Probabilistic Inference: Learning Meets Sampling}, Singapore, April 2025{\natexlab{a}}.
\newblock \doi{10.48550/arXiv.2502.06079}.
\newblock URL \url{https://iclr.cc/virtual/2025/workshop/23990}.
\newblock Oral; non-archival.

\bibitem[Lee et~al.(2025{\natexlab{b}})Lee, Park, Chu, Yoon, and Kim]{lee_latent_2025}
Seunghun Lee, Jinyoung Park, Jaewon Chu, Minseo Yoon, and Hyunwoo~J. Kim.
\newblock Latent bayesian optimization via autoregressive normalizing flows.
\newblock In \emph{Proceedings of the Thirteenth International Conference on Learning Representations (ICLR)}, Singapore, April 2025{\natexlab{b}}.
\newblock URL \url{https://openreview.net/forum?id=ZCOwwRAaEl}.
\newblock ICLR 2025 (Oral).

\bibitem[Li et~al.(2025{\natexlab{a}})Li, Yang, Johnston, G{\"u}rsoy, Yue, and Arnold]{li_evaluation_2024}
Francesca-Zhoufan Li, Jason Yang, Kadina~E. Johnston, Emre G{\"u}rsoy, Yisong Yue, and Frances~H. Arnold.
\newblock Evaluation of machine learning-assisted directed evolution across diverse combinatorial landscapes.
\newblock \emph{Cell Systems}, 16\penalty0 (9):\penalty0 101387, September 2025{\natexlab{a}}.
\newblock \doi{10.1016/j.cels.2025.101387}.
\newblock URL \url{https://doi.org/10.1016/j.cels.2025.101387}.

\bibitem[Li et~al.(2025{\natexlab{b}})Li, Chen, Guo, and Shen]{li_survey_2025}
Tianyi Li, Mingda Chen, Bowei Guo, and Zhiqiang Shen.
\newblock A survey on diffusion language models.
\newblock \emph{arXiv preprint arXiv:2508.10875}, August 2025{\natexlab{b}}.
\newblock \doi{10.48550/arXiv.2508.10875}.
\newblock URL \url{https://arxiv.org/abs/2508.10875}.
\newblock Preprint.

\bibitem[Li et~al.(2022)Li, Thickstun, Gulrajani, Liang, and Hashimoto]{li_diffusion-lm_2022}
Xiang~Lisa Li, John Thickstun, Ishaan Gulrajani, Percy Liang, and Tatsunori~B. Hashimoto.
\newblock Diffusion-{LM} improves controllable text generation.
\newblock In \emph{Advances in Neural Information Processing Systems (NeurIPS)}, volume~35, 2022.
\newblock URL \url{https://proceedings.neurips.cc/paper_files/paper/2022/hash/1be5bc25d50895ee656b8c2d9eb89d6a-Abstract-Conference.html}.
\newblock NeurIPS 2022.

\bibitem[Li et~al.(2024{\natexlab{a}})Li, Zhao, Wang, Scalia, Eraslan, Nair, Biancalani, Ji, Regev, Levine, and Uehara]{li_derivative-free_2024}
Xiner Li, Yulai Zhao, Chenyu Wang, Gabriele Scalia, G{\"o}kcen Eraslan, Surag Nair, Tommaso Biancalani, Shuiwang Ji, Aviv Regev, Sergey Levine, and Masatoshi Uehara.
\newblock Derivative-free guidance in continuous and discrete diffusion models with soft value-based decoding.
\newblock In \emph{NeurIPS 2024 Workshop on AI for New Drug Modalities}, New Orleans, LA, USA, December 2024{\natexlab{a}}.
\newblock URL \url{https://neurips.cc/virtual/2024/102888}.
\newblock Workshop poster; non-archival.

\bibitem[Li et~al.(2024{\natexlab{b}})Li, Yuan, Huang, Ni, Ye, Chen, and Wang]{li_diffusion_2024}
Zihao Li, Hui Yuan, Kaixuan Huang, Chengzhuo Ni, Yinyu Ye, Minshuo Chen, and Mengdi Wang.
\newblock Diffusion model for data-driven black-box optimization.
\newblock \emph{arXiv preprint arXiv:2403.13219}, March 2024{\natexlab{b}}.
\newblock \doi{10.48550/arXiv.2403.13219}.
\newblock URL \url{https://arxiv.org/abs/2403.13219}.
\newblock Preprint.

\bibitem[Lim et~al.(2025)Lim, Lee, and No]{lim_scoring-assisted_2025}
Hocheol Lim, Geon-Ho Lee, and Kyoung~Tai No.
\newblock Scoring-assisted generative exploration for proteins (sage-prot): A framework for multi-objective protein optimization via iterative sequence generation and evaluation.
\newblock \emph{arXiv preprint arXiv:2505.01277}, May 2025.
\newblock \doi{10.48550/arXiv.2505.01277}.
\newblock URL \url{https://arxiv.org/abs/2505.01277}.
\newblock Preprint.

\bibitem[Lin et~al.(2023)Lin, Akin, Rao, Hie, Zhu, Lu, Smetanin, Verkuil, Kabeli, Shmueli, dos Santos~Costa, Fazel-Zarandi, Sercu, Candido, and Rives]{lin_evolutionary-scale_2023}
Zeming Lin, Halil Akin, Roshan Rao, Brian Hie, Zhongkai Zhu, Wenting Lu, Nikita Smetanin, Robert Verkuil, Ori Kabeli, Yaniv Shmueli, Allan dos Santos~Costa, Maryam Fazel-Zarandi, Tom Sercu, Salvatore Candido, and Alexander Rives.
\newblock Evolutionary-scale prediction of atomic-level protein structure with a language model.
\newblock \emph{Science}, 379\penalty0 (6637):\penalty0 1123--1130, March 2023.
\newblock \doi{10.1126/science.ade2574}.
\newblock URL \url{https://www.science.org/doi/10.1126/science.ade2574}.

\bibitem[Lisanza et~al.(2025)Lisanza, Gershon, Tipps, Sims, Arnoldt, Hendel, Simma, Liu, Yase, Wu, Tharp, Li, Kang, Brackenbrough, Bera, Gerben, Wittmann, McShan, and Baker]{lisanza_multistate_2024}
Sidney~Lyayuga Lisanza, Jacob~Merle Gershon, Samuel W.~K. Tipps, Jeremiah~Nelson Sims, Lucas Arnoldt, Samuel~J. Hendel, Miriam~K. Simma, Ge~Liu, Muna Yase, Hongwei Wu, Claire~D. Tharp, Xinting Li, Alex Kang, Evans Brackenbrough, Asim~K. Bera, Stacey Gerben, Bruce~J. Wittmann, Andrew~C. McShan, and David Baker.
\newblock Multistate and functional protein design using {RoseTTAFold} sequence space diffusion.
\newblock \emph{Nature Biotechnology}, 43\penalty0 (8):\penalty0 1288--1298, August 2025.
\newblock \doi{10.1038/s41587-024-02395-w}.
\newblock URL \url{https://www.nature.com/articles/s41587-024-02395-w}.
\newblock Epub 2024-09-25; Publisher Correction: Nat Biotechnol 43(8):1384, doi:10.1038/s41587-024-02456-0.

\bibitem[Liu et~al.(2025)Liu, Nam, Campbell, St{\"a}rk, Xu, Jaakkola, and G{\'o}mez-Bombarelli]{liu_think_2024}
Sulin Liu, Juno Nam, Andrew Campbell, Hannes St{\"a}rk, Yilun Xu, Tommi Jaakkola, and Rafael G{\'o}mez-Bombarelli.
\newblock Think while you generate: Discrete diffusion with planned denoising.
\newblock In \emph{Proceedings of the Thirteenth International Conference on Learning Representations (ICLR)}, Singapore, April 2025.
\newblock URL \url{https://openreview.net/forum?id=MJNywBdSDy}.
\newblock Poster.

\bibitem[Lou et~al.(2024)Lou, Meng, and Ermon]{lou_discrete_2024}
Aaron Lou, Chenlin Meng, and Stefano Ermon.
\newblock Discrete diffusion modeling by estimating the ratios of the data distribution.
\newblock In \emph{Proceedings of the 41st International Conference on Machine Learning (ICML)}, volume 235 of \emph{Proceedings of Machine Learning Research}, pp.\  32819--32848, Vienna, Austria, 21--27 Jul 2024. PMLR.
\newblock URL \url{https://proceedings.mlr.press/v235/lou24a.html}.

\bibitem[Mardani et~al.(2024)Mardani, Song, Kautz, and Vahdat]{mardani2024a}
Morteza Mardani, Jiaming Song, Jan Kautz, and Arash Vahdat.
\newblock A variational perspective on solving inverse problems with diffusion models.
\newblock In \emph{The Twelfth International Conference on Learning Representations}, 2024.
\newblock URL \url{https://openreview.net/forum?id=1YO4EE3SPB}.

\bibitem[Maus et~al.(2022)Maus, Jones, Moore, Kusner, Bradshaw, and Gardner]{maus2022local}
Natalie Maus, Haydn Jones, Juston Moore, Matt~J Kusner, John Bradshaw, and Jacob Gardner.
\newblock Local latent space bayesian optimization over structured inputs.
\newblock In S.~Koyejo, S.~Mohamed, A.~Agarwal, D.~Belgrave, K.~Cho, and A.~Oh (eds.), \emph{Advances in Neural Information Processing Systems}, volume~35, pp.\  34505--34518. Curran Associates, Inc., 2022.
\newblock URL \url{https://proceedings.neurips.cc/paper_files/paper/2022/file/ded98d28f82342a39f371c013dfb3058-Paper-Conference.pdf}.

\bibitem[Meshchaninov et~al.(2025)Meshchaninov, Strashnov, Shevtsov, Nikolaev, Ivanisenko, Kardymon, and Vetrov]{meshchaninov_diffusion_2025}
Viacheslav Meshchaninov, Pavel Strashnov, Andrey Shevtsov, Fedor Nikolaev, Nikita Ivanisenko, Olga Kardymon, and Dmitry Vetrov.
\newblock Diffusion on language model encodings for protein sequence generation.
\newblock In \emph{Proceedings of the 42nd International Conference on Machine Learning (ICML)}, Vancouver, Canada, July 2025.
\newblock \doi{10.48550/arXiv.2403.03726}.
\newblock URL \url{https://icml.cc/virtual/2025/poster/43588}.
\newblock Poster; to appear.

\bibitem[Michael et~al.(2024)Michael, Kæstel-Hansen, Mørch~Groth, Bartels, Salomon, Tian, Hatzakis, and Boomsma]{michael_systematic_2024}
Richard Michael, Jacob Kæstel-Hansen, Peter Mørch~Groth, Simon Bartels, Jesper Salomon, Pengfei Tian, Nikos~S. Hatzakis, and Wouter Boomsma.
\newblock A systematic analysis of regression models for protein engineering.
\newblock \emph{PLOS Computational Biology}, 20\penalty0 (5):\penalty0 e1012061, May 2024.
\newblock ISSN 1553-7358.
\newblock \doi{10.1371/journal.pcbi.1012061}.
\newblock URL \url{https://dx.plos.org/10.1371/journal.pcbi.1012061}.

\bibitem[Nichol \& Dhariwal(2021)Nichol and Dhariwal]{pmlr-v139-nichol21a}
Alexander~Quinn Nichol and Prafulla Dhariwal.
\newblock Improved denoising diffusion probabilistic models.
\newblock In Marina Meila and Tong Zhang (eds.), \emph{Proceedings of the 38th International Conference on Machine Learning}, volume 139 of \emph{Proceedings of Machine Learning Research}, pp.\  8162--8171. PMLR, 18--24 Jul 2021.
\newblock URL \url{https://proceedings.mlr.press/v139/nichol21a.html}.

\bibitem[Nie et~al.(2025)Nie, Zhu, You, Zhang, Ou, Hu, Zhou, Lin, Wen, and Li]{nie_large_2025}
Shen Nie, Fengqi Zhu, Zebin You, Xiaolu Zhang, Jingyang Ou, Jun Hu, Jun Zhou, Yankai Lin, Ji-Rong Wen, and Chongxuan Li.
\newblock Large language diffusion models.
\newblock In \emph{Advances in Neural Information Processing Systems (NeurIPS)}, San Diego, CA, USA, December 2025.
\newblock URL \url{https://neurips.cc/virtual/2025/poster/118608}.

\bibitem[Nijkamp et~al.(2023)Nijkamp, Ruffolo, Weinstein, Naik, and Madani]{nijkamp_progen2_2023}
Erik Nijkamp, Jeffrey~A. Ruffolo, Eli~N. Weinstein, Nikhil Naik, and Ali Madani.
\newblock {ProGen2}: {Exploring} the boundaries of protein language models.
\newblock \emph{Cell Systems}, 14\penalty0 (11):\penalty0 968--978.e3, November 2023.
\newblock ISSN 24054712.
\newblock \doi{10.1016/j.cels.2023.10.002}.
\newblock URL \url{https://linkinghub.elsevier.com/retrieve/pii/S2405471223002727}.

\bibitem[Nisonoff et~al.(2025)Nisonoff, Xiong, Allenspach, and Listgarten]{nisonoff_unlocking_2024}
Hunter Nisonoff, Junhao Xiong, Stephan Allenspach, and Jennifer Listgarten.
\newblock Unlocking guidance for discrete state-space diffusion and flow models.
\newblock In \emph{Proceedings of the Thirteenth International Conference on Learning Representations (ICLR)}, Singapore, April 2025.
\newblock URL \url{https://openreview.net/forum?id=XsgHl54yO7}.
\newblock ICLR 2025 — Poster.

\bibitem[Olson et~al.(2014)Olson, Wu, and Sun]{olson_comprehensive_2014}
C.~Anders Olson, Nicholas~C. Wu, and Ren Sun.
\newblock A {Comprehensive} {Biophysical} {Description} of {Pairwise} {Epistasis} throughout an {Entire} {Protein} {Domain}.
\newblock \emph{Current Biology}, 24\penalty0 (22):\penalty0 2643--2651, November 2014.
\newblock ISSN 09609822.
\newblock \doi{10.1016/j.cub.2014.09.072}.
\newblock URL \url{https://linkinghub.elsevier.com/retrieve/pii/S0960982214012688}.

\bibitem[Packer \& Liu(2015)Packer and Liu]{packer_methods_2015}
Michael~S. Packer and David~R. Liu.
\newblock Methods for the directed evolution of proteins.
\newblock \emph{Nature Reviews Genetics}, 16\penalty0 (7):\penalty0 379--394, July 2015.
\newblock ISSN 1471-0056, 1471-0064.
\newblock \doi{10.1038/nrg3927}.
\newblock URL \url{http://www.nature.com/articles/nrg3927}.

\bibitem[Peng et~al.(2025)Peng, Bezemek, Patel, Rector-Brooks, Yao, Tong, and Chatterjee]{peng_path_2025}
Fred~Zhangzhi Peng, Zachary Bezemek, Sawan Patel, Jarrid Rector-Brooks, Sherwood Yao, Alexander Tong, and Pranam Chatterjee.
\newblock Path planning for masked diffusion model sampling.
\newblock In \emph{ICLR 2025 DeLTa Workshop}, Singapore, April 2025.
\newblock \doi{10.48550/arXiv.2502.03540}.
\newblock URL \url{https://openreview.net/forum?id=fFuVPKpSt0}.
\newblock Workshop poster; non-archival.

\bibitem[Rafailov et~al.(2023)Rafailov, Sharma, Mitchell, Ermon, Manning, and Finn]{rafailov_direct_2023}
Rafael Rafailov, Archit Sharma, Eric Mitchell, Stefano Ermon, Christopher~D. Manning, and Chelsea Finn.
\newblock Direct preference optimization: Your language model is secretly a reward model.
\newblock In \emph{Advances in Neural Information Processing Systems (NeurIPS)}, volume~36, 2023.
\newblock URL \url{https://papers.nips.cc/paper_files/paper/2023/hash/a85b405ed65c6477a4fe8302b5e06ce7-Abstract-Conference.html}.
\newblock NeurIPS 2023.

\bibitem[Rector-Brooks et~al.(2025)Rector-Brooks, Hasan, Peng, Quinn, Liu, Mittal, Dziri, Bronstein, Bengio, Chatterjee, Tong, and Bose]{rector-brooks_steering_2024}
Jarrid Rector-Brooks, Mohsin Hasan, Zhangzhi Peng, Zachary Quinn, Chenghao Liu, Sarthak Mittal, Nouha Dziri, Michael Bronstein, Yoshua Bengio, Pranam Chatterjee, Alexander Tong, and Avishek~Joey Bose.
\newblock Steering masked discrete diffusion models via discrete denoising posterior prediction.
\newblock In \emph{Proceedings of the Thirteenth International Conference on Learning Representations (ICLR)}, Singapore, April 2025.
\newblock \doi{10.48550/arXiv.2410.08134}.
\newblock URL \url{https://openreview.net/forum?id=Ombm8S40zN}.
\newblock Poster.

\bibitem[Ren et~al.(2022)Ren, Li, Ding, Zhou, Ma, and Peng]{pmlr-v162-ren22a}
Zhizhou Ren, Jiahan Li, Fan Ding, Yuan Zhou, Jianzhu Ma, and Jian Peng.
\newblock Proximal exploration for model-guided protein sequence design.
\newblock In Kamalika Chaudhuri, Stefanie Jegelka, Le~Song, Csaba Szepesvari, Gang Niu, and Sivan Sabato (eds.), \emph{Proceedings of the 39th International Conference on Machine Learning}, volume 162 of \emph{Proceedings of Machine Learning Research}, pp.\  18520--18536. PMLR, 17--23 Jul 2022.
\newblock URL \url{https://proceedings.mlr.press/v162/ren22a.html}.

\bibitem[Rives et~al.(2021)Rives, Meier, Sercu, Goyal, Lin, Liu, Guo, Ott, Zitnick, Ma, and Fergus]{rives_biological_2021}
Alexander Rives, Joshua Meier, Tom Sercu, Siddharth Goyal, Zeming Lin, Jason Liu, Demi Guo, Myle Ott, C.~Lawrence Zitnick, Jerry Ma, and Rob Fergus.
\newblock Biological structure and function emerge from scaling unsupervised learning to 250 million protein sequences.
\newblock \emph{Proceedings of the National Academy of Sciences}, 118\penalty0 (15), April 2021.
\newblock ISSN 0027-8424, 1091-6490.
\newblock \doi{10.1073/pnas.2016239118}.
\newblock URL \url{https://www.pnas.org/content/118/15/e2016239118}.
\newblock Publisher: National Academy of Sciences Section: Biological Sciences.

\bibitem[Romero \& Arnold(2009)Romero and Arnold]{romero_exploring_2009}
Philip~A Romero and Frances~H Arnold.
\newblock Exploring protein fitness landscapes by directed evolution.
\newblock \emph{Nat Rev Mol Cell Biol}, 10:\penalty0 866--876, 2009.
\newblock \doi{10.1038/nrm2805}.

\bibitem[Ruffolo \& Madani(2024)Ruffolo and Madani]{ruffolo_designing_2024}
Jeffrey~A Ruffolo and Ali Madani.
\newblock Designing proteins with language models.
\newblock \emph{Nature Biotechnology}, 42:\penalty0 200--202, February 2024.

\bibitem[Russo et~al.(2018)Russo, Van~Roy, Kazerouni, Osband, Wen, et~al.]{russo2018tutorial}
Daniel~J Russo, Benjamin Van~Roy, Abbas Kazerouni, Ian Osband, Zheng Wen, et~al.
\newblock A tutorial on thompson sampling.
\newblock \emph{Foundations and Trends{\textregistered} in Machine Learning}, 11\penalty0 (1):\penalty0 1--96, 2018.

\bibitem[Sahoo et~al.(2024)Sahoo, Arriola, Schiff, Gokaslan, Marroquin, Chiu, Rush, and Kuleshov]{sahoo_simple_2024}
Subham~Sekhar Sahoo, Marianne Arriola, Yair Schiff, Aaron Gokaslan, Edgar Marroquin, Justin~T. Chiu, Alexander Rush, and Volodymyr Kuleshov.
\newblock Simple and effective masked diffusion language models.
\newblock In \emph{Advances in Neural Information Processing Systems (NeurIPS)}, volume~37 of \emph{Advances in Neural Information Processing Systems}, Vancouver, Canada, December 2024.
\newblock URL \url{https://proceedings.neurips.cc/paper_files/paper/2024/hash/eb0b13cc515724ab8015bc978fdde0ad-Abstract-Conference.html}.

\bibitem[Schiff et~al.(2024)Schiff, Sahoo, Phung, Wang, Boshar, Dalla-torre, Almeida, Rush, Pierrot, and Kuleshov]{schiff_simple_2024}
Yair Schiff, Subham~Sekhar Sahoo, Hao Phung, Guanghan Wang, Sam Boshar, Hugo Dalla-torre, Bernardo P.~de Almeida, Alexander Rush, Thomas Pierrot, and Volodymyr Kuleshov.
\newblock Simple {Guidance} {Mechanisms} for {Discrete} {Diffusion} {Models}, December 2024.
\newblock URL \url{http://arxiv.org/abs/2412.10193}.
\newblock arXiv:2412.10193 [cs].

\bibitem[Seki et~al.(2025)Seki, Guo, Akpinaroglu, and Kortemme]{seki_combinatorial_2025}
Kosuke Seki, Amy~B. Guo, Deniz Akpinaroglu, and Tanja Kortemme.
\newblock A combinatorial mutational map of active non-native protein kinases by deep learning guided sequence design.
\newblock \emph{bioRxiv}, pp.\  2025.08.03.668353, August 2025.
\newblock \doi{10.1101/2025.08.03.668353}.
\newblock URL \url{https://www.biorxiv.org/content/10.1101/2025.08.03.668353}.
\newblock Preprint; version 1 posted Aug 3, 2025.

\bibitem[Shi et~al.(2024)Shi, Han, Wang, Doucet, and Titsias]{shi2025simplifiedgeneralizedmaskeddiffusion}
Jiaxin Shi, Kehang Han, Zhe Wang, Arnaud Doucet, and Michalis~K. Titsias.
\newblock Simplified and generalized masked diffusion for discrete data.
\newblock 2024.
\newblock URL \url{https://arxiv.org/abs/2406.04329}.

\bibitem[Sinai et~al.(2020)Sinai, Wang, Whatley, Slocum, Locane, and Kelsic]{sinai2020adaleadsimplerobustadaptive}
Sam Sinai, Richard Wang, Alexander Whatley, Stewart Slocum, Elina Locane, and Eric~D. Kelsic.
\newblock Adalead: A simple and robust adaptive greedy search algorithm for sequence design, 2020.
\newblock URL \url{https://arxiv.org/abs/2010.02141}.

\bibitem[Singhal et~al.(2025)Singhal, Horvitz, Teehan, Ren, Yu, McKeown, and Ranganath]{singhal_general_2025}
Raghav Singhal, Zachary Horvitz, Ryan Teehan, Mengye Ren, Zhou Yu, Kathleen McKeown, and Rajesh Ranganath.
\newblock A general framework for inference-time scaling and steering of diffusion models.
\newblock In \emph{Proceedings of the 42nd International Conference on Machine Learning (ICML)}, Vancouver, Canada, July 2025.
\newblock \doi{10.48550/arXiv.2501.06848}.
\newblock URL \url{https://icml.cc/virtual/2025/poster/45673}.
\newblock ICML 2025 (poster).

\bibitem[Soares et~al.(2025)Soares, Hetzel, Szymczak, Theis, G{\"u}nnemann, and Szczurek]{soares_targeted_2025}
Diogo Soares, Leon Hetzel, Paulina Szymczak, Fabian Theis, Stephan G{\"u}nnemann, and Ewa Szczurek.
\newblock Targeted {AMP} generation through controlled diffusion with efficient embeddings, April 2025.
\newblock URL \url{https://arxiv.org/abs/2504.17247}.
\newblock Preprint.

\bibitem[Song et~al.(2021)Song, Sohl-Dickstein, Kingma, Kumar, Ermon, and Poole]{song2020score}
Yang Song, Jascha Sohl-Dickstein, Diederik~P. Kingma, Abhishek Kumar, Stefano Ermon, and Ben Poole.
\newblock Score-based generative modeling through stochastic differential equations.
\newblock In \emph{Proceedings of the International Conference on Learning Representations (ICLR)}, May 2021.
\newblock URL \url{https://openreview.net/forum?id=PxTIG12RRHS}.
\newblock ICLR 2021 — Outstanding Paper Award.

\bibitem[Song \& Li(2023)Song and Li]{song_importance_2024}
Zhenqiao Song and Lei Li.
\newblock Importance weighted expectation-maximization for protein sequence design.
\newblock In Andreas Krause, Emma Brunskill, Kyunghyun Cho, Barbara Engelhardt, Sivan Sabato, and Jonathan Scarlett (eds.), \emph{Proceedings of the 40th International Conference on Machine Learning}, volume 202 of \emph{Proceedings of Machine Learning Research}, pp.\  32349--32364, Honolulu, Hawaii, USA, Jul 2023. PMLR.
\newblock URL \url{https://proceedings.mlr.press/v202/song23g.html}.
\newblock ICML 2023.

\bibitem[Stanton et~al.(2022)Stanton, Maddox, Gruver, Maffettone, Delaney, Greenside, and Wilson]{stanton_accelerating_2022}
Samuel Stanton, Wesley Maddox, Nate Gruver, Phillip Maffettone, Emily Delaney, Peyton Greenside, and Andrew~Gordon Wilson.
\newblock Accelerating {Bayesian} {Optimization} for {Biological} {Sequence} {Design} with {Denoising} {Autoencoders}.
\newblock In \emph{Proceedings of the 39th International Conference on Machine Learning}, volume 162 of \emph{Proceedings of Machine Learning Research}, pp.\  20459--20478. PMLR, 17--23 Jul 2022.
\newblock URL \url{https://proceedings.mlr.press/v162/stanton22a.html}.

\bibitem[Stark et~al.(2024)Stark, Jing, Wang, Corso, Berger, Barzilay, and Jaakkola]{stark_dirichlet_2024}
Hannes Stark, Bowen Jing, Chenyu Wang, Gabriele Corso, Bonnie Berger, Regina Barzilay, and Tommi Jaakkola.
\newblock Dirichlet flow matching with applications to {DNA} sequence design.
\newblock In \emph{Proceedings of the 41st International Conference on Machine Learning (ICML)}, volume 235 of \emph{Proceedings of Machine Learning Research}, pp.\  46495--46513, Vienna, Austria, 21--27 Jul 2024. PMLR.
\newblock URL \url{https://proceedings.mlr.press/v235/stark24b.html}.

\bibitem[Steinegger \& Söding(2017)Steinegger and Söding]{steinegger_mmseqs2_2017}
Martin Steinegger and Johannes Söding.
\newblock {MMseqs2} enables sensitive protein sequence searching for the analysis of massive data sets.
\newblock \emph{Nature Biotechnology}, 35\penalty0 (11):\penalty0 1026--1028, November 2017.
\newblock ISSN 1546-1696.
\newblock \doi{10.1038/nbt.3988}.
\newblock URL \url{https://www.nature.com/articles/nbt.3988}.
\newblock Number: 11 Publisher: Nature Publishing Group.

\bibitem[Stocco et~al.(2024)Stocco, Artigues-Lleix{\`a}, Hunklinger, Widatalla, G{\"u}ell, and Ferruz]{stocco_guiding_nodate}
Filippo Stocco, Maria Artigues-Lleix{\`a}, Andrea Hunklinger, Talal Widatalla, Marc G{\"u}ell, and Noelia Ferruz.
\newblock Guiding generative protein language models with reinforcement learning.
\newblock In \emph{NeurIPS 2024 Workshop on Machine Learning in Structural Biology (MLSB)}, Vancouver, Canada, December 2024.
\newblock \doi{10.48550/arXiv.2412.12979}.
\newblock URL \url{https://slideslive.com/39031185/guiding-protein-language-models-with-reinforcement-learning}.
\newblock Workshop presentation (non-archival).

\bibitem[Sumida et~al.(2024)Sumida, Núñez-Franco, Kalvet, Pellock, Wicky, Milles, Dauparas, Wang, Kipnis, Jameson, Kang, De~La~Cruz, Sankaran, Bera, Jiménez-Osés, and Baker]{sumida_improving_2024}
Kiera~H. Sumida, Reyes Núñez-Franco, Indrek Kalvet, Samuel~J. Pellock, Basile I.~M. Wicky, Lukas~F. Milles, Justas Dauparas, Jue Wang, Yakov Kipnis, Noel Jameson, Alex Kang, Joshmyn De~La~Cruz, Banumathi Sankaran, Asim~K. Bera, Gonzalo Jiménez-Osés, and David Baker.
\newblock Improving {Protein} {Expression}, {Stability}, and {Function} with {ProteinMPNN}.
\newblock \emph{Journal of the American Chemical Society}, 146\penalty0 (3):\penalty0 2054--2061, January 2024.
\newblock ISSN 0002-7863, 1520-5126.
\newblock \doi{10.1021/jacs.3c10941}.
\newblock URL \url{https://pubs.acs.org/doi/10.1021/jacs.3c10941}.

\bibitem[Sun et~al.(2025)Sun, He, Deng, Liu, Zhao, Jiang, Cao, Ju, Wu, Liu, Qin, and Liu]{sun_accelerating_2025}
Haoran Sun, Liang He, Pan Deng, Guoqing Liu, Zhiyu Zhao, Yuliang Jiang, Chuan Cao, Fusong Ju, Lijun Wu, Haiguang Liu, Tao Qin, and Tie-Yan Liu.
\newblock Accelerating protein engineering with fitness landscape modelling and reinforcement learning.
\newblock \emph{Nature Machine Intelligence}, September 2025.
\newblock ISSN 2522-5839.
\newblock \doi{10.1038/s42256-025-01103-w}.
\newblock URL \url{https://doi.org/10.1038/s42256-025-01103-w}.

\bibitem[Tang et~al.(2025{\natexlab{a}})Tang, Zhang, and Chatterjee]{tang_peptune_2024}
Sophia Tang, Yinuo Zhang, and Pranam Chatterjee.
\newblock Peptune: De novo generation of therapeutic peptides with multi-objective-guided discrete diffusion.
\newblock In \emph{Proceedings of the 42nd International Conference on Machine Learning (ICML 2025)}, Vancouver, Canada, July 2025{\natexlab{a}}.
\newblock \doi{10.48550/arXiv.2412.17780}.
\newblock URL \url{https://icml.cc/virtual/2025/poster/45889}.
\newblock Poster; ICML 2025.

\bibitem[Tang et~al.(2025{\natexlab{b}})Tang, Zhang, Tong, and Chatterjee]{tang_gumbel-softmax_2025}
Sophia Tang, Yinuo Zhang, Alexander Tong, and Pranam Chatterjee.
\newblock Gumbel-softmax flow matching with straight-through guidance for controllable biological sequence generation.
\newblock In \emph{ICLR 2025 Workshop on Integrating Generative and Experimental Platforms for Biomolecular Design (GEM)}, 2025{\natexlab{b}}.
\newblock URL \url{https://arxiv.org/abs/2503.17361}.
\newblock Workshop poster (non-archival); also available as arXiv:2503.17361.

\bibitem[Thomas et~al.(2025)Thomas, Belanger, Xu, Lee, Hirano, Iwai, Polic, Nyberg, Hoff, Frenz, Emrich, Kim, Chavarha, Ramanan, Agresti, and Colwell]{thomas_engineering_2025}
Neil Thomas, David Belanger, Chenling Xu, Hanson Lee, Kathleen Hirano, Kosuke Iwai, Vanja Polic, Kendra~D. Nyberg, Kevin~G. Hoff, Lucas Frenz, Charlie~A. Emrich, Jun~W. Kim, Mariya Chavarha, Abi Ramanan, Jeremy~J. Agresti, and Lucy~J. Colwell.
\newblock Engineering highly active nuclease enzymes with machine learning and high-throughput screening.
\newblock \emph{Cell Systems}, 16\penalty0 (3):\penalty0 101236, March 2025.
\newblock ISSN 2405-4712.
\newblock \doi{10.1016/j.cels.2025.101236}.
\newblock URL \url{https://www.sciencedirect.com/science/article/pii/S2405471225000699}.

\bibitem[Torres et~al.(2024)Torres, Zeng, Wan, Maus, Gardner, and de~la Fuente-Nunez]{torres_generative_2024}
Marcelo D.~T. Torres, Yimeng Zeng, Fangping Wan, Natalie Maus, Jacob Gardner, and Cesar de~la Fuente-Nunez.
\newblock A generative artificial intelligence approach for antibiotic optimization.
\newblock \emph{bioRxiv}, November 2024.
\newblock \doi{10.1101/2024.11.27.625757}.
\newblock URL \url{https://www.biorxiv.org/content/10.1101/2024.11.27.625757v1}.
\newblock preprint.

\bibitem[Torres et~al.(2025)Torres, Chen, Wan, Chatterjee, and de~la Fuente]{torres_generative_2025}
Marcelo D.~T. Torres, Tianlai Chen, Fangping Wan, Pranam Chatterjee, and Cesar de~la Fuente.
\newblock Generative latent diffusion language modeling yields anti-infective synthetic peptides.
\newblock \emph{Cell Biomaterials}, 1:\penalty0 100183, October 2025.
\newblock \doi{10.1016/j.celbio.2025.100183}.
\newblock URL \url{https://www.cell.com/cell-biomaterials/fulltext/S3050-5623(25)00174-6}.
\newblock Published version of bioRxiv preprint 10.1101/2025.01.31.636003; online early Sept 2, 2025.

\bibitem[Uehara et~al.(2025)Uehara, Su, Zhao, Li, Regev, Ji, Levine, and Biancalani]{uehara_reward-guided_2025}
Masatoshi Uehara, Xingyu Su, Yulai Zhao, Xiner Li, Aviv Regev, Shuiwang Ji, Sergey Levine, and Tommaso Biancalani.
\newblock Reward-guided iterative refinement in diffusion models at test-time with applications to protein and dna design.
\newblock In \emph{Proceedings of the 42nd International Conference on Machine Learning (ICML)}, Proceedings of Machine Learning Research, Vancouver, Canada, July 2025. PMLR.
\newblock URL \url{https://icml.cc/virtual/2025/poster/46204}.
\newblock ICML 2025 (poster).

\bibitem[Venkatraman et~al.(2024)Venkatraman, Jain, Scimeca, Kim, Sendera, Hasan, Rowe, Mittal, Lemos, Bengio, Adam, Rector-Brooks, Bengio, Berseth, and Malkin]{venkatraman_amortizing_2025}
Siddarth Venkatraman, Moksh Jain, Luca Scimeca, Minsu Kim, Marcin Sendera, Mohsin Hasan, Luke Rowe, Sarthak Mittal, Pablo Lemos, Emmanuel Bengio, Alexandre Adam, Jarrid Rector-Brooks, Yoshua Bengio, Glen Berseth, and Nikolay Malkin.
\newblock Amortizing intractable inference in diffusion models for vision, language, and control.
\newblock In \emph{Advances in Neural Information Processing Systems 37 (NeurIPS 2024)}, Vancouver, Canada, 2024.
\newblock URL \url{https://neurips.cc/virtual/2024/poster/95348}.
\newblock NeurIPS 2024.

\bibitem[Venkatraman et~al.(2025)Venkatraman, Hasan, Kim, Scimeca, Sendera, Bengio, Berseth, and Malkin]{venkatraman_outsourced_2025}
Siddarth Venkatraman, Mohsin Hasan, Minsu Kim, Luca Scimeca, Marcin Sendera, Yoshua Bengio, Glen Berseth, and Nikolay Malkin.
\newblock Outsourced diffusion sampling: Efficient posterior inference in latent spaces of generative models.
\newblock In \emph{Proceedings of the 42nd International Conference on Machine Learning}, volume 267 of \emph{Proceedings of Machine Learning Research}, pp.\  1--28, Vancouver, BC, Canada, July 2025. PMLR.
\newblock \doi{10.48550/arXiv.2502.06999}.
\newblock URL \url{https://openreview.net/forum?id=94c9hu6Fsv}.
\newblock ICML 2025 (poster). Also available as arXiv:2502.06999.

\bibitem[Vornholt et~al.(2024)Vornholt, Mutný, Schmidt, Schellhaas, Tachibana, Panke, Ward, Krause, and Jeschek]{vornholt_enhanced_2024}
Tobias Vornholt, Mojmír Mutný, Gregor~W. Schmidt, Christian Schellhaas, Ryo Tachibana, Sven Panke, Thomas~R. Ward, Andreas Krause, and Markus Jeschek.
\newblock Enhanced {Sequence}-{Activity} {Mapping} and {Evolution} of {Artificial} {Metalloenzymes} by {Active} {Learning}.
\newblock \emph{ACS Central Science}, 10\penalty0 (7):\penalty0 1357--1370, May 2024.
\newblock ISSN 2374-7943, 2374-7951.
\newblock URL \url{https://pubs.acs.org/doi/10.1021/acscentsci.4c00258}.

\bibitem[Wang et~al.(2025{\natexlab{a}})Wang, Uehara, He, Wang, Biancalani, Lal, Jaakkola, Levine, Wang, and Regev]{wang_fine-tuning_2024}
Chenyu Wang, Masatoshi Uehara, Yichun He, Amy Wang, Tommaso Biancalani, Avantika Lal, Tommi Jaakkola, Sergey Levine, Hanchen Wang, and Aviv Regev.
\newblock Fine-tuning discrete diffusion models via reward optimization with applications to dna and protein design.
\newblock In \emph{Proceedings of the Thirteenth International Conference on Learning Representations (ICLR 2025)}, Singapore, April 2025{\natexlab{a}}.
\newblock URL \url{https://openreview.net/forum?id=G328D1xt4W}.
\newblock ICLR 2025 Poster.

\bibitem[Wang et~al.(2025{\natexlab{b}})Wang, Schiff, Sahoo, and Kuleshov]{wang_remasking_2025}
Guanghan Wang, Yair Schiff, Subham~Sekhar Sahoo, and Volodymyr Kuleshov.
\newblock Remasking discrete diffusion models with inference-time scaling.
\newblock In \emph{Advances in Neural Information Processing Systems 38 (NeurIPS 2025)}, December 2025{\natexlab{b}}.
\newblock URL \url{https://neurips.cc/virtual/2025/poster/118818}.
\newblock Poster.

\bibitem[Wang et~al.(2024)Wang, Zheng, Ye, Xue, Huang, and Gu]{wang_diffusion_2024}
Xinyou Wang, Zaixiang Zheng, Fei Ye, Dongyu Xue, Shujian Huang, and Quanquan Gu.
\newblock Diffusion language models are versatile protein learners.
\newblock In \emph{Proceedings of the 41st International Conference on Machine Learning}, volume 235 of \emph{Proceedings of Machine Learning Research}, pp.\  52309--52333, Vienna, Austria, Jul 2024. PMLR.
\newblock URL \url{https://proceedings.mlr.press/v235/wang24ct.html}.

\bibitem[Wang et~al.(2025{\natexlab{c}})Wang, Fan, Guo, Nguyen, Ji, and Liu]{wang_proteinzero_2025}
Ziwen Wang, Jiajun Fan, Ruihan Guo, Thao Nguyen, Heng Ji, and Ge~Liu.
\newblock Proteinzero: Self-improving protein generation via online reinforcement learning, jun 2025{\natexlab{c}}.
\newblock URL \url{https://arxiv.org/abs/2506.07459}.
\newblock arXiv preprint.

\bibitem[Widatalla et~al.(2024)Widatalla, Rafailov, and Hie]{widatalla_aligning_2024}
Talal Widatalla, Rafael Rafailov, and Brian Hie.
\newblock Aligning protein generative models with experimental fitness via {Direct Preference Optimization}.
\newblock \emph{bioRxiv}, 2024.
\newblock \doi{10.1101/2024.05.20.595026}.
\newblock URL \url{https://doi.org/10.1101/2024.05.20.595026}.

\bibitem[Wilson et~al.(2016)Wilson, Hu, Salakhutdinov, and Xing]{wilson_deep_2015}
Andrew~Gordon Wilson, Zhiting Hu, Ruslan Salakhutdinov, and Eric~P. Xing.
\newblock Deep kernel learning.
\newblock In \emph{Proceedings of the 19th International Conference on Artificial Intelligence and Statistics}, volume~51 of \emph{Proceedings of Machine Learning Research}, pp.\  370--378, C{\'a}diz, Spain, 2016. PMLR.
\newblock URL \url{https://proceedings.mlr.press/v51/}.

\bibitem[Wittmann et~al.(2024)Wittmann, Alexanian, Bartling, Beal, Clore, Diggans, Flyangolts, Gemler, Mitchell, Murphy, Wheeler, and Horvitz]{wittmann_toward_2024}
Bruce Wittmann, Tessa Alexanian, Craig Bartling, Jacob Beal, Adam Clore, James Diggans, Kevin Flyangolts, Bryan~T. Gemler, Tom Mitchell, Steven~T. Murphy, Nicole~E. Wheeler, and Eric Horvitz.
\newblock Toward {AI}-{Resilient} {Screening} of {Nucleic} {Acid} {Synthesis} {Orders}: {Process}, {Results}, and {Recommendations}.
\newblock \emph{bioRxiv}, December 2024.
\newblock \doi{10.1101/2024.12.02.626439}.
\newblock URL \url{https://www.biorxiv.org/content/10.1101/2024.12.02.626439}.
\newblock Preprint.

\bibitem[Wittmann et~al.(2021{\natexlab{a}})Wittmann, Johnston, Wu, and Arnold]{wittmann_advances_2021}
Bruce~J. Wittmann, Kadina~E. Johnston, Zachary Wu, and Frances~H. Arnold.
\newblock Advances in machine learning for directed evolution.
\newblock \emph{Current Opinion in Structural Biology}, 69:\penalty0 11--18, 2021{\natexlab{a}}.
\newblock ISSN 1879033X.
\newblock \doi{10.1016/j.sbi.2021.01.008}.
\newblock URL \url{https://doi.org/10.1016/j.sbi.2021.01.008}.
\newblock Publisher: Elsevier Ltd.

\bibitem[Wittmann et~al.(2021{\natexlab{b}})Wittmann, Yue, and Arnold]{wittmann_informed_2021}
Bruce~J. Wittmann, Yisong Yue, and Frances~H. Arnold.
\newblock Informed training set design enables efficient machine learning-assisted directed protein evolution.
\newblock \emph{Cell Systems}, 12\penalty0 (11):\penalty0 1026--1045.e7, 2021{\natexlab{b}}.
\newblock ISSN 24054712.
\newblock \doi{10.1016/j.cels.2021.07.008}.
\newblock URL \url{https://doi.org/10.1016/j.cels.2021.07.008}.
\newblock Publisher: Elsevier Inc.

\bibitem[Wu et~al.(2025)Wu, Kuang, Niu, Ma, and Yu]{wu_diffusion-bbo_2025}
Dongxia Wu, Nikki~Lijing Kuang, Ruijia Niu, Yi-An Ma, and Rose Yu.
\newblock Diffusion-{BBO}: {Diffusion}-{Based} {Inverse} {Modeling} for {Online} {Black}-{Box} {Optimization}, 2025.
\newblock URL \url{https://arxiv.org/abs/2407.00610}.
\newblock arXiv preprint; also presented as a poster at NeurIPS 2024 Workshop on Bayesian Decision-making and Uncertainty.

\bibitem[Wu et~al.(2024)Wu, Trippe, Naesseth, Blei, and Cunningham]{wu2024practical}
Luhuan Wu, Brian~L. Trippe, Christian~A. Naesseth, David~M. Blei, and John~P. Cunningham.
\newblock Practical and asymptotically exact conditional sampling in diffusion models, 2024.
\newblock URL \url{https://arxiv.org/abs/2306.17775}.

\bibitem[Wu et~al.(2019)Wu, Kan, Lewis, Wittmann, and Arnold]{wu_machine_2019}
Zachary Wu, S.~B.~Jennifer Kan, Russell~D. Lewis, Bruce~J. Wittmann, and Frances~H. Arnold.
\newblock Machine learning-assisted directed protein evolution with combinatorial libraries.
\newblock \emph{Proceedings of the National Academy of Sciences}, 116\penalty0 (18):\penalty0 8852--8858, April 2019.
\newblock ISSN 0027-8424, 1091-6490.
\newblock \doi{10.1073/pnas.1901979116}.
\newblock URL \url{http://www.pnas.org/lookup/doi/10.1073/pnas.1901979116}.

\bibitem[Wu et~al.(2021)Wu, Johnston, Arnold, and Yang]{wu_protein_2021}
Zachary Wu, Kadina~E. Johnston, Frances~H. Arnold, and Kevin~K. Yang.
\newblock Protein sequence design with deep generative models.
\newblock \emph{Current Opinion in Chemical Biology}, 65:\penalty0 18--27, 2021.
\newblock ISSN 13675931.
\newblock \doi{10.1016/j.cbpa.2021.04.004}.
\newblock URL \url{http://arxiv.org/abs/2104.04457%0Ahttp://dx.doi.org/10.1016/j.cbpa.2021.04.004}.
\newblock arXiv: 2104.04457 Publisher: Elsevier Ltd.

\bibitem[Xiong et~al.(2025)Xiong, Nisonoff, Gaur, Lukarska, Oltrogge, Savage, and Listgarten]{xiong_guide_2025}
Junhao Xiong, Hunter Nisonoff, Ishan Gaur, Maria~E. Lukarska, Luke~M. Oltrogge, David~F. Savage, and Jennifer Listgarten.
\newblock Guide your favorite protein sequence generative model, May 2025.
\newblock URL \url{https://arxiv.org/abs/2505.04823}.
\newblock arXiv:2505.04823.

\bibitem[Yang et~al.(2024)Yang, Li, and Arnold]{yang_opportunities_2024}
Jason Yang, Francesca-Zhoufan Li, and Frances~H. Arnold.
\newblock Opportunities and {Challenges} for {Machine} {Learning}-{Assisted} {Enzyme} {Engineering}.
\newblock \emph{ACS Central Science}, 10\penalty0 (2):\penalty0 226--241, February 2024.
\newblock ISSN 2374-7943, 2374-7951.
\newblock \doi{10.1021/acscentsci.3c01275}.
\newblock URL \url{https://pubs.acs.org/doi/10.1021/acscentsci.3c01275}.

\bibitem[Yang et~al.(2025{\natexlab{a}})Yang, Bhatnagar, Ruffolo, and Madani]{yang_conditional_2024}
Jason Yang, Aadyot Bhatnagar, Jeffrey~A. Ruffolo, and Ali Madani.
\newblock Function-guided conditional generation using protein language models with adapters.
\newblock \emph{arXiv}, June 2025{\natexlab{a}}.
\newblock \doi{10.48550/arXiv.2410.03634}.
\newblock URL \url{https://arxiv.org/abs/2410.03634}.
\newblock arXiv:2410.03634 [q-bio.BM], v2, last revised 2025-06-11.

\bibitem[Yang et~al.(2025{\natexlab{b}})Yang, Lal, Bowden, Astudillo, Hameedi, Kaur, Hill, Yue, and Arnold]{yang_active_2025}
Jason Yang, Ravi~G. Lal, James~C. Bowden, Raul Astudillo, Mikhail~A. Hameedi, Sukhvinder Kaur, Matthew Hill, Yisong Yue, and Frances~H. Arnold.
\newblock Active learning-assisted directed evolution.
\newblock \emph{Nature Communications}, 16\penalty0 (1):\penalty0 714, January 2025{\natexlab{b}}.
\newblock ISSN 2041-1723.
\newblock \doi{10.1038/s41467-025-55987-8}.
\newblock URL \url{https://www.nature.com/articles/s41467-025-55987-8}.
\newblock Publisher: Nature Publishing Group.

\bibitem[Yang et~al.(2025{\natexlab{c}})Yang, Li, Long, and Arnold]{yang_illuminating_2025}
Jason Yang, Francesca-Zhoufan Li, Yueming Long, and Frances~H. Arnold.
\newblock Illuminating the universe of enzyme catalysis in the era of artificial intelligence.
\newblock \emph{Cell Systems}, pp.\  101372, August 2025{\natexlab{c}}.
\newblock ISSN 2405-4712.
\newblock \doi{10.1016/j.cels.2025.101372}.
\newblock URL \url{https://www.sciencedirect.com/science/article/pii/S2405471225002054}.

\bibitem[Yang \& Klein(2021)Yang and Klein]{yang_fudge_2021}
Kevin Yang and Dan Klein.
\newblock {FUDGE}: {Controlled} {Text} {Generation} {With} {Future} {Discriminators}.
\newblock In \emph{Proceedings of the 2021 {Conference} of the {North} {American} {Chapter} of the {Association} for {Computational} {Linguistics}: {Human} {Language} {Technologies}}, pp.\  3511--3535, 2021.
\newblock \doi{10.18653/v1/2021.naacl-main.276}.
\newblock URL \url{http://arxiv.org/abs/2104.05218}.
\newblock arXiv:2104.05218 [cs].

\bibitem[Yang et~al.(2019)Yang, Wu, and Arnold]{yang_machine-learning-guided_2019}
Kevin~K. Yang, Zachary Wu, and Frances~H. Arnold.
\newblock Machine-learning-guided directed evolution for protein engineering.
\newblock \emph{Nature Methods}, 16\penalty0 (8):\penalty0 687--694, 2019.
\newblock ISSN 15487105.
\newblock \doi{10.1038/s41592-019-0496-6}.
\newblock URL \url{http://dx.doi.org/10.1038/s41592-019-0496-6}.
\newblock arXiv: 1811.10775 Publisher: Springer US.

\bibitem[Zhang et~al.(2025)Zhang, Chu, Berner, Meng, Anandkumar, and Song]{zhang_improving_2024}
Bingliang Zhang, Wenda Chu, Julius Berner, Chenlin Meng, Anima Anandkumar, and Yang Song.
\newblock Improving diffusion inverse problem solving with decoupled noise annealing.
\newblock In \emph{Proceedings of the IEEE/CVF Conference on Computer Vision and Pattern Recognition (CVPR)}. IEEE, June 2025.
\newblock \doi{10.1109/CVPR52734.2025.01946}.
\newblock URL \url{https://arxiv.org/abs/2407.01521}.
\newblock CVPR 2025; arXiv:2407.01521.

\bibitem[Zhao et~al.(2024{\natexlab{a}})Zhao, Zhang, and Luo]{zhao_contrastive_2024}
Junming Zhao, Chao Zhang, and Yunan Luo.
\newblock Contrastive {Fitness} {Learning}: {Reprogramming} {Protein} {Language} {Models} for {Low}-{N} {Learning} of {Protein} {Fitness} {Landscape}.
\newblock preprint, Bioinformatics, February 2024{\natexlab{a}}.
\newblock URL \url{http://biorxiv.org/lookup/doi/10.1101/2024.02.11.579859}.

\bibitem[Zhao et~al.(2024{\natexlab{b}})Zhao, Brekelmans, Makhzani, and Grosse]{zhao_probabilistic_2024}
Stephen Zhao, Rob Brekelmans, Alireza Makhzani, and Roger~Baker Grosse.
\newblock Probabilistic {Inference} in {Language} {Models} via {Twisted} {Sequential} {Monte} {Carlo}.
\newblock In Ruslan Salakhutdinov, Zico Kolter, Katherine Heller, Adrian Weller, Nuria Oliver, Jonathan Scarlett, and Felix Berkenkamp (eds.), \emph{Proceedings of the 41st International Conference on Machine Learning}, volume 235 of \emph{Proceedings of Machine Learning Research}, pp.\  60704--60748. PMLR, July 2024{\natexlab{b}}.
\newblock URL \url{https://proceedings.mlr.press/v235/zhao24c.html}.

\bibitem[Zheng et~al.(2025)Zheng, Chu, Zhang, Wu, Wang, Feng, Zou, Sun, Kovachki, Ross, Bouman, and Yue]{zheng2025inversebench}
Hongkai Zheng, Wenda Chu, Bingliang Zhang, Zihui Wu, Austin Wang, Berthy~T. Feng, Caifeng Zou, Yu~Sun, Nikola~Borislavov Kovachki, Zachary~E. Ross, Katherine~L. Bouman, and Yisong Yue.
\newblock Inversebench: Benchmarking plug-and-play diffusion priors for inverse problems in physical sciences.
\newblock In \emph{Proceedings of the Thirteenth International Conference on Learning Representations (ICLR 2025)}, Singapore, may 2025. OpenReview.
\newblock URL \url{https://openreview.net/forum?id=U3PBITXNG6}.
\newblock ICLR 2025 Spotlight.

\end{thebibliography}
\bibliographystyle{iclr2025_conference}

\newpage
\appendix
\section{Appendix}

\setcounter{figure}{0}
\setcounter{table}{0}  
\renewcommand{\thefigure}{A\arabic{figure}}  
\renewcommand{\theHfigure}{\thesection.\arabic{figure}}
\renewcommand{\thetable}{A\arabic{table}}        
\renewcommand{\theHtable}{\thesection.\arabic{table}}

\subsection{Data for pretraining generative priors}
\label{section:pretraining_data}

The first step in our pipeline involves learning a generative prior on naturally occurring protein sequences to capture the distribution of those with high evolutionary likelihood. This prior is unconditional in the sense that no labeled fitness data is used for training. However, because we are optimizing protein \textit{variants} for a desired fitness, we pretrained our generative prior on sequences homologous to the parent protein to be optimized (known as a multiple sequence alignment or MSA): TrpB, CreiLOV, or GB1. Likelihoods from MSAs have been captured by statistical models and have been shown to offer good zero-shot approximations of fitness. In other words, they capture mutational substitutions that are more favorable, based on the precedent of natural evolution.

We focused on  the TrpB \citep{johnston_combinatorially_2024} and CreiLOV \citep{chen_deep_2023} datasets due to the extensive number of sequences in their MSAs and compared to GB1 \citep{olson_comprehensive_2014}, which has comparatively fewer sequences. MSAs were obtained by running jackhmmer \citep{johnson_hidden_2010} against Uniref90 for two iterations with the parent sequence of the fitness dataset as target. For the MSA, we only used sequences where the aligned portion was at least 75\% the length of the parent sequence. We used the MSA that was aligned to the parent sequence, with gap tokens replaced by the corresponding amino acid found in the parent sequence, resulting in full, fixed-length pseudo-natural sequences. For GB1, we augmented the training set with synthetic data, namely all proteins with a single mutation to sequences in the MSA. For the language models on TrpB and CreiLOV, some sequences were randomly mutated by a single position near the beginning of the sequence, to prevent mode collapse during autoregressive generation.

We performed sequence clustering using mmseqs2 \citep{steinegger_mmseqs2_2017} at 80\% identity and resampled the dataset by weighting each sample with $\frac{1}{1 + \ln{(n)}}$ relative probability of being sampled, where $n$ is the size of the cluster associated with that sequence. Afterward, we removed 5\% of the clusters and their associated sequences as a validation set.

\subsection{Protein fitness optimization task}
\label{section:task}

\paragraph{An oracle as a proxy for protein fitness.}
We studied fitness optimization across three different protein-fitness datasets, TrpB, CreiLOV, and GB1 (Table \ref{table:data}). TrpB is 389 residues in length, but based on available fitness data, we limited design to 15 residues: 117, 118, 119, 162, 166, 182, 183, 184, 185, 186, 227, 228, 230, 231, and 301. Namely, we combined the fitness data from 6 combinatorially complete 3-site libraries (D-I from \citet{johnston_combinatorially_2024}) and the 4-site library across residues 183, 184, 227, and 228. We normalized the parent fitness to 1 in each dataset and rounded all negative fitness values up to zero. The fitness here is the catalytic rate of a native reaction, the formation of tryptophan from indole and serine. To obtain a proxy fitness for all variants in the design space ($20^{15}$ possibilities) we trained an oracle inspired by the dataset splitting and model architecture used in  \citet{blalock_functional_nodate}. Namely, we used all of the single, double, and triple mutants in the library for training, with 10\% and 20\% of the quadruple mutants being used for validation and testing, respectively. Our model consists of an ensemble of 20 MLPs for TrpB, and each was trained on one-hot encodings of the designed residues for 1000 epochs. 

Differently, the CreiLOV dataset (length $N = 119$) contains experimental fitnesses for all single mutations in the protein and certain higher order mutations at 15 selected positions with beneficial single mutations. Fitness here refers to associated fluorescence. To obtain a proxy fitness for all variants in the design space ($20^{119}$ possibilities), we trained an oracle similar to the procedure above, using similar splits to those in \citet{blalock_functional_nodate} and were able to reproduce their high performance on the test set. Before model training, we scaled the fitnesses of the single mutants to the fitnesses of multi-mutants by adding a normalization factor to all single mutants such that the parent sequence in both datasets had the same fitness.  Our model consists of an ensemble of 10 MLPs for CreiLOV, and each was trained on onehot encodings of sequences for 1000 epochs.

For GB1, the experimental finesses for nearly all double mutations across the entire protein were available, where fitness refers to binding affinity of a domain of the G protein. To train the oracle, we held out 10\% and 20\% the sequences with two mutations as a validation and test set, respectively, with remaining sequences being used for training. Our model consists of an ensemble of 10 MLPs for GB1, and each was trained on one-hot encodings of the designed residues for 50 epochs.

\begin{figure}[h]
  \centering
\includegraphics[width=\textwidth]{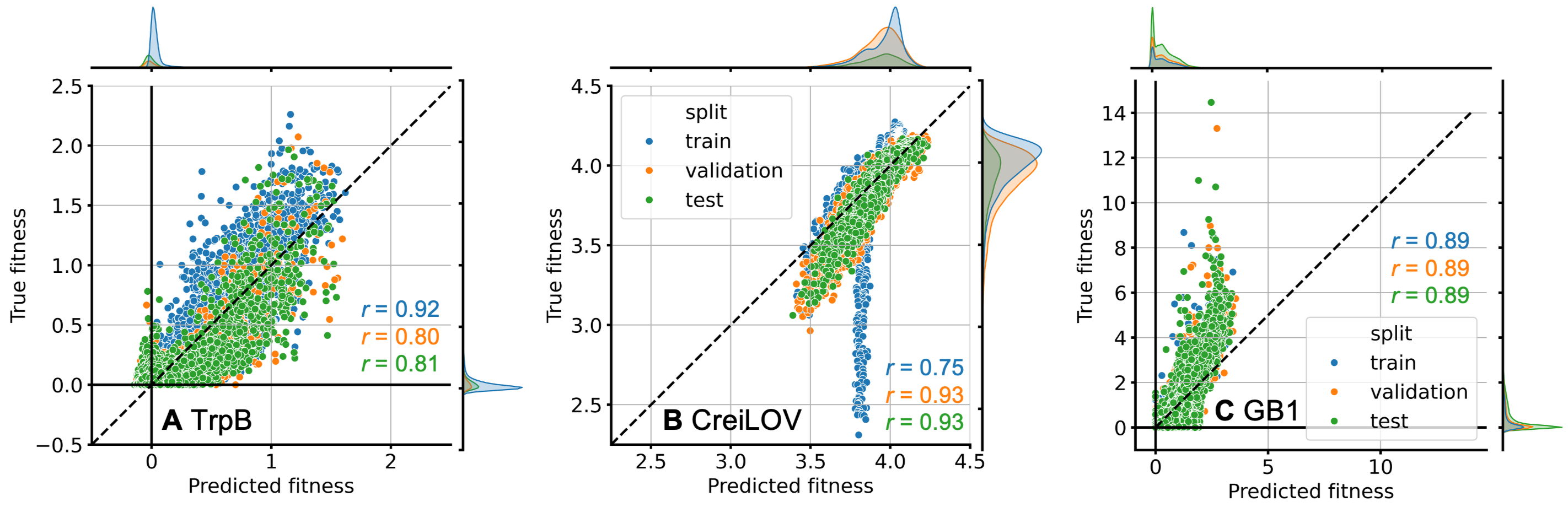}
  \vspace{-0.1in}
  \caption{Oracles trained on available labeled fitness data for TrpB, CreiLOV, and GB1 extrapolate well to higher order combinations of mutations within the design space, as measured by Pearson correlation.}
  \vspace{-0.1in}
  \label{fig:oracle}
\end{figure}

 Our oracles show high Pearson correlation on the train and test sets (Fig. \ref{fig:oracle}). As the generalization ability of our oracle was only been tested on variants that are similar to the parent, we penalized the fitness of protein sequences by a factor of 0.99 for every mutation accumulated beyond a threshold of 60\% sequence identity to the parent sequence. From here forth, we treated ground truth fitness as outputs from the oracle. 

\begin{figure}[h]
  \centering
\includegraphics[width=\textwidth]{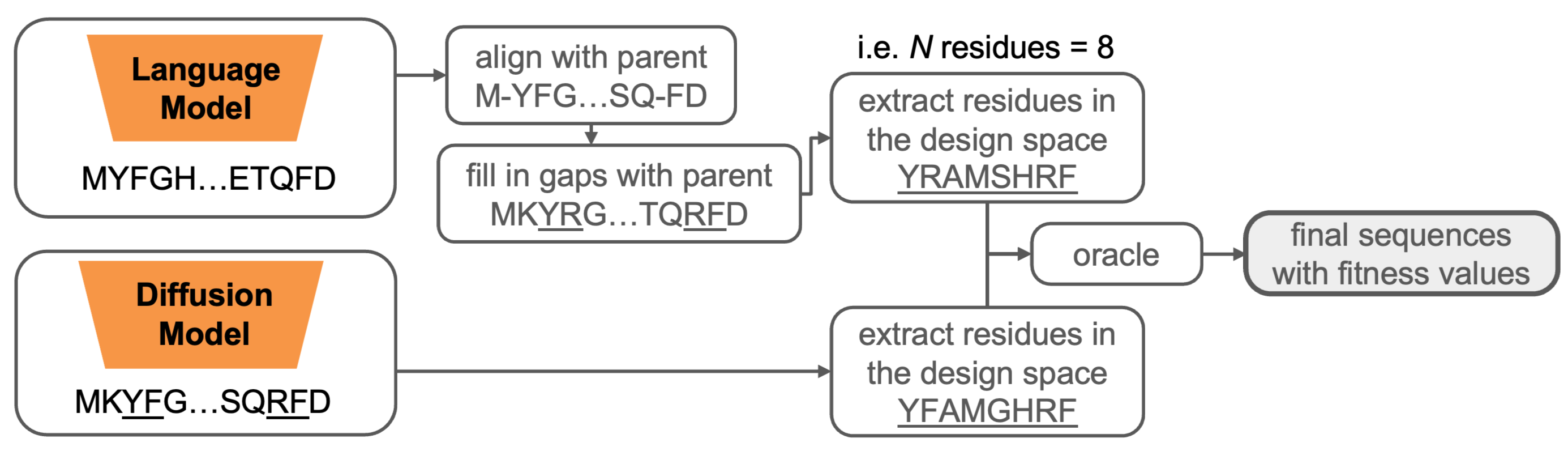}
  \vspace{-0.3in}
  \caption{Example pipeline for generating protein sequences for evaluation, based on a hypothetical parent sequence: MKKFG...SQRFD (length=100), with 8 residues being optimized (3, 4, 26, 27, 28, 29, 98, 99), corresponding to a design space combo of KFDEACRF.}
  \vspace{-0.1in}
  \label{fig:generation_pipeline}
\end{figure}

\paragraph{Processing generated sequences.}
Our primary method for evaluation involved examining the distribution of sampled sequences and their corresponding fitness values, diversities, and novelties. The processing pipeline for generated sequences in shown in Fig. \ref{fig:generation_pipeline}. In diffusion models, sequences were generated with fixed length equal to the parent length. For the language models, nearly all generated sequences had length equal to the parent sequence length. Still, sequences were aligned with the parent sequence using mafft \citep{katoh_mafft_2013}, and gaps were replaced with the corresponding amino acid in the parent sequence to generate complete pseudo-sequences of a fixed length. Special tokens, which occurred rarely in generation, were replaced by a random amino acid. For TrpB, residues outside of the design space of 15 residues were naively mapped to the original amino acid type in the parent sequence at the end of generation. We did not test inpainting, although this could be accomplished with masked (diffusion) language models.

 \paragraph{Comparison to existing protein engineering methods.} 
There are several reasons why we did not directly compare the performance of SGPO methods to existing methods used in protein engineering, such as directed evolution and MLDE. In the case of directed evolution (such as random mutagenesis): (1) It is not obvious which parent sequences to use as the starting points for directed evolution for a fair comparison. (2) It is unclear if the oracle captures the true nature of the protein fitness landscape or extrapolates well to sequences with many mutations relative to the original fitness dataset from which the oracle was trained. (3) Overall, our method enables the accumulation of many mutations in a single round of experimentation, whereas directed evolution is largely limited to one mutation at a time. For example, on the CreiLOV dataset, the generated sequences with the highest fitness had on average 66 mutations from the parent reference sequence from which the original dataset was generated, which would not be achievable with directed evolution. We also did not directly compare our method to supervised approaches in smaller design spaces, such as 4-site combinatorial libraries \citep{yang_active_2025}, as we focus here on design in larger design spaces, where existing methods are lacking.
Overall, traversing large swaths of sequence space will be important for faster engineering and enabling improvements to fitness that would normally be slow with directed evolution.
 
\subsection{Generative models for sequences}
\label{section:model_methods}

\begin{table}[h]
  \caption{\textbf{Summary of training details for generative priors in this work.} Reference refers to the codebase that was modified for our implementation and where the model architecture was adapted from. For all models, we retained the model with the lowest validation loss. When using the ESM encoder, we used the 35M-parameter ESM2 model \cite{lin_evolutionary-scale_2023}.}
  \centering
  \begin{tabular}{p{1.4cm}  p{0.7cm} p{1.2cm}  p{0.5cm} p{0.9cm} p{1.3cm} p{1.2cm} p{1.5cm} p{1.8cm}} 
    \\
    \toprule
    Model  & Max Epochs & Learning Rate & Batch Size & Warmup Steps & Noise Schedule & Diffusion Timesteps & Model \newline Architecture & Reference \\
    \midrule
    \textbf{Continuous} & 5 & $1\times10^{-4}$ & 64 & 10 & cosine & 500 & BERT & \citet{gruver_protein_2023} \\
    Continuous-ESM & 25 & $1\times10^{-4}$& 64 & 10 & cosine & 500 & BERT & \citet{gruver_protein_2023} \\
    \midrule
     D3PM-Baseline & 5 & $1\times10^{-4}$ & 64 & 10 & Sohl-Dickstein & 500 & ByteNet & \citet{alamdari_protein_2023} \\
   \textbf{ D3PM} & 5 & $1\times10^{-4}$ & 64 & 10 & Sohl-Dickstein & 500 & ByteNet  & \citet{alamdari_protein_2023} \\
   UDLM & 5 & $3\times10^{-5}$ & 64 & 2500 & loglinear & 500 & DiT & \citet{schiff_simple_2024}\\
   \midrule
   \textbf{ MDLM} & 50 & $3\times10^{-4}$ & 64 & 2500 & loglinear & 500 & DiT & \citet{schiff_simple_2024} \\
    \midrule
   \textbf{ ARLM} & 10 & $1\times10^{-4}$ & 32 & 10 & n/a & n/a & GPT-J & \citet{nijkamp_progen2_2023}\\
    \bottomrule
  \end{tabular}
  \label{table:additional_training}
\end{table}

\subsubsection{Diffusion over continuous space}
Diffusion models construct samples by reversing a diffusion process that maps clean data points $\rvx_0$ to samples from a prior distribution $\pi(\rvx)$. The forward process $(\rvx_0 \to \rvx_T)$ is composed of conditional distributions $p(\rvx_t|\rvx_{t-1})$, which admit closed-form expressions for the conditional distributions $p(\rvx_t|\rvx_0)$ and $p(\rvx_{t-1}|\rvx_t, \rvx_0)$. The reverse process $(\rvx_T \to \rvx_0)$ converts samples from the prior into samples from the learned data distribution $p_\theta(\rvx_0)$ by repeatedly predicting the denoised variable $\hat{\rvx}_0$ from noisy values $\rvx_t$, using the conditional distribution $p(\rvx_{t-1}|\rvx_t, \hat{\rvx}_0)$ to derive a transition distribution $p_\theta(\rvx_{t-1}|\rvx_t)$.

\paragraph{Continuous noise forward process.}
Similarly to \cite{gruver_protein_2023}, we define a protein sequence as \( \rvw \in \mathcal{A}^{L} \), where \( \mathcal{A} \) is the alphabet of amino acids and \( L \) is the fixed length of the sequence. To learn a distribution $p(\rvw)$, we first embed $\rvw$ into a continuous variable $\rvx_0$ using an embedding matrix $U_\theta$ or encoder from the ESM2 language model \citep{lin_evolutionary-scale_2023}, transforming discrete tokens into a continuous latent space. Gaussian noise is then applied to this embedding space. The prior distribution is defined as:
\begin{equation}
    \pi(\rvx) = \mathcal{N}(0, I),
\end{equation}
while the forward process follows a Gaussian corruption schedule:
\begin{equation}
    p(\rvx_t | \rvx_0) = \mathcal{N}(\sqrt{\bar{\alpha}_t} \rvx_0, (1 - \bar{\alpha}_t)I), \quad \bar{\alpha}_t = \prod_{i=1}^{t} \alpha_i, \quad \alpha_t = 1 - \beta_t.
\end{equation}
The variance schedule $\{\beta_t\}$ follows the cosine schedule proposed by \cite{pmlr-v139-nichol21a}, which is commonly used to stabilize training.

\paragraph{Reverse process.}
The reverse process aims to recover the original sequence by learning a function $p_\theta(\hat{\rvw} | \rvx_t,t)$ that predicts the sequence from noised points $\rvx_t$. This is done by minimizing the following objective:
\begin{equation}
    L(\theta) = \mathbb{E}_{\rvw_0,t} \left[ - \log p_\theta(\rvw_0 | \rvx_t) \right], \quad \rvx_t \sim p(\rvx_t | \rvx_0 = U_\theta \rvw_0).
\end{equation}
By learning $p_\theta(\hat{\rvw} | \rvx_t, t)$, we construct the reverse transition distribution:
\begin{equation}
    p_\theta(\rvx_{t-1} | \rvx_t) = \sum_{\hat{\rvw}} p(\rvx_{t-1} | \rvx_t, \hat{\rvx}_0 = U_\theta \hat{\rvw}) p_\theta(\hat{\rvw} | \rvx_t, t),
\end{equation}
where the posterior $p(\rvx_{t-1} | \rvx_t, \rvx_0)$ follows:
\begin{equation}
    p(\rvx_{t-1} | \rvx_t, \rvx_0) = \mathcal{N}(\rvx_{t-1}; \mu_t, \sigma_t^2 I),
\end{equation}
with mean $\mu_t$ and variance $\sigma_t^2$ given by:
\begin{equation}
    \mu_t = \frac{\sqrt{\bar{\alpha}_{t-1}} \beta_t}{1 - \bar{\alpha}_t} \rvx_0 + \frac{\sqrt{\alpha_t} (1 - \bar{\alpha}_{t-1})}{1 - \bar{\alpha}_t} \rvx_t,
\end{equation}
\begin{equation}
    \sigma_t^2 = \frac{1 - \bar{\alpha}_{t-1}}{1 - \bar{\alpha}_t} \beta_t.
\end{equation}

\paragraph{Inference and sampling.}
At inference time, the learned reverse process is used to generate protein sequences from the prior $\pi(x)$. This is done by iteratively sampling:
\begin{equation}
    \rvx_{t-1} \sim p_\theta(\rvx_{t-1} | \rvx_t),
\end{equation}
and then reconstructing $\rvw$ by sampling:
\begin{equation}
    \rvw \sim p_\theta(\hat{\rvw} | \rvx_0).
\end{equation}
This denoising process iteratively refines noisy embeddings back into structured sequences.


\subsubsection{Diffusion over discrete space.}
Discrete diffusion models \citep{austin_structured_2023,campbell_continuous_2022,lou_discrete_2024} generate data in discrete spaces by reversing a predefined forward Markov process. Specifically, a family of distributions $p_t$ evolves according to the Markov chain
\begin{equation}
    \frac{\mathrm dp_t}{\mathrm dt} = \bm Q_t p_t,
\end{equation}
where $p_0 = p_{\mathrm{data}}$ is the data distribution and $\bm Q_t\in \mathbb R^{N\times N}$ are predefined transition matrices.

This Markov process can be reversed with the help of a concrete score function, $s(\rvx,t) := [\frac{p_t(\tilde \rvx)}{p_t(\rvx)}]_{\tilde \rvx \neq \rvx}$, as its time reversal is given by
\begin{equation}
    \frac{\mathrm dp_{T-t}}{\mathrm dt} = \bar {\bm Q}_{T-t} p_{T-t}, \label{eq:reverse_ctmc}
\end{equation}
where $\bar {\bm Q}_{t}[\tilde \rvx,\rvx] = s(\rvx,t)_{\tilde x} \bm Q_t [\rvx,\tilde \rvx]$ for $\tilde \rvx \neq \rvx$, and $\bar {\bm Q}_t[\rvx,\rvx] = - \sum_{\tilde \rvx \neq \rvx} \bar {\bm Q}_t[\tilde \rvx, \rvx]$. To generate data $\rvx_0 \sim p_{\mathrm{data}}$, we start with sampling $\rvx_T$ from a uniform distribution and then evolve through Eq.~\ref{eq:reverse_ctmc} by the Euler method.

\paragraph{Uniform discrete language models.} Both D3PM~\citep{austin_structured_2023} and UDLM~\citep{schiff_simple_2024} implement a uniform transition matrix $\bm Q_t = \frac{1}{N} \mathbf 1\mathbf 1^T - \bm I$. When $T\to \infty$, the probability distribution $p_T$ converges to a uniform distribution.

\paragraph{Masked diffusion language models.} Masked diffusion language models (MDLM)~\citep{sahoo_simple_2024} utilize an absorbing transition matrix $\mQ_t$ that converts tokens in a sequence to [MASK] states. The corresponding transition matrix can be written as $\bm Q_t\in \mathbb R^{(N+1)\times (N+1)}$, $\mQ_t = -\mI + \mathbf e_{N+1}\mathbf 1^T$. When $T\to \infty$, the limiting distribution $p_T$ converges to a completely masked sequence.


\subsubsection{Autoregressive language models.} In this work, we finetuned the ProGen2-small decoder-only transformer (151 million parameters) based on the code and parameters used in \citet{yang_conditional_2024}. Models were trained based on next token prediction and cross entropy loss. However, we did not use adapter layers, and we did not group batches based on sequence length. During inference from the autoregressive model, we used a temperature of 1 and a Top-\textit{p} value of 1. 

\subsection{Steering methods}
\label{section:steering_methods}
\begin{table}[h]
  \caption{\textbf{Summary of supervised value functions used to predict fitness in this work, to guide diffusion models.} All “classifiers'' were trained as regressors to predict fitness. For DAPS methods, only clean data was used for training, whereas other classifiers are trained on clean and noised samples from various timesteps.}
  \centering
  \begin{tabular}{p{1.4cm}  p{1.4cm} p{1.2cm}  p{1.5cm} p{1.2cm} p{1.8cm} p{1.8cm}} 
    \\
    \toprule
    Model  & Guidance Strategy & Max Epochs & Learning Rate & Batch Size & Architecture & Hidden Dimension \\
    \midrule
    Continuous Diffusion & CG & 1000 & $1\times10^{-3}$ & 128 & 4-layer MLP & 256 \\
     & DAPS & 1000 & $1\times10^{-3}$ & 128 & 4-layer MLP & 256 \\
    \midrule
    Continuous Diffusion & NOS & 100 & $1\times10^{-3}$& 128 & 1-layer MLP & 256 \\
    \midrule
     Discrete Diffusion & CG & 1000 & $1\times10^{-3}$ & 64 & 4-layer MLP & 64 \\
    & DAPS & 200 & $1\times10^{-3}$& 64 & 4-layer MLP  & 64 \\
    \midrule
    Discrete Diffusion & NOS & 200 & $3\times10^{-4}$& 64 & linear layer & n/a  \\
    \bottomrule
  \end{tabular}
  \label{table:classifier_training}
\end{table}

\begin{table}[h]
  \caption{\textbf{Hyperparameters used to tune the guidance/steering process.} The \textbf{bolded} parameter was chosen as the ideal parameter for the iterative ``Bayesian optimization'' experiment (Fig. \ref{fig:iterative}). Larger guidance parameter typically implements stronger guidance strength.}
  \label{table:guidance_parameters}
  \centering
  \begin{tabular}{p{2.8cm}  p{7cm}} 
    \\
    \toprule
    Guidance Strategy  & Hyperparameters \\
    \midrule
    Continuous CG & $1/\beta = 64, 128, 256, 512, 1024$ \\
    Discrete CG  & $1/\beta = 1, 2.5, 6.25, \textbf{15.625}, 39.0625$  \\
    \midrule
    Continuous DAPS & $1/\beta = 0.25, 0.5,1, 2, 4 \times 10^4$ \\ 
    & $K = 50$ \\ 
    & Euler method steps = 10\\
    & Langevin dynamics steps = 100\\
    Discrete DAPS & $1/\beta = 16, 32, 64, \textbf{128}, 256$ \\ 
    & $K = 50$ \\ 
    & Euler method steps = 20\\
    & Metropolis Hastings steps = 1000\\
    \midrule
    Continuous NOS &
    $\lambda = 0.1, 1, 10, 100, \textbf{1000}$  \\
     & $\eta = 0.5, \textbf{2}, 5 $  \\
     & $K = 5, \textbf{10} $  \\
     & optimizer = AdaGrad  \\
    Discrete NOS & 
    $\lambda = 0.1, 1, 10, 100, 1000$  \\
     & $\eta = 0.5, 2, 5 $  \\
     & $K = 5, 10 $  \\
     & optimizer = AdaGrad  \\
    \midrule
    DPO & $\beta = 0.02, 0.1, 0.5, \textbf{2}, 4 $\\
    & lr = $1 \times 10^{-6}$ \\ 
    & epochs = 5 \\
    & batch size = 8 \\
    \bottomrule
  \end{tabular}
  
\end{table}

\subsubsection{Classifier guidance}
Classifier guidance \citep{song2020score} is a technique used to steer samples generated by diffusion models toward desired attributes. The primary goal is to sample from a conditional distribution $p(\rvx|\rvy)$, where $\rvy$ is a guiding signal of interest. In continuous space, this can be achieved by replacing the unconditional score function $\nabla_{\rvx_t} \log p_t(\rvx_t)$ at time $t$ by a conditional score function,
\begin{equation}
    \nabla_{\rvx_t} \log p(\rvx_t|\rvy) = \nabla_{\rvx_t} \log p_t(\rvx_t) + \nabla_{\rvx_t} \log p_t(\rvy|\rvx_t) 
    \label{eq:conditional_score}
\end{equation}
To obtain the conditional score function, one only needs to train a time-dependent predictor, which predicts the probability of $p_t(\rvy|\rvx_t)$ given $\rvx_t$ and time $t$.

\paragraph{Continuous guidance.}
Classifier guidance modifies the reverse diffusion process to steer generated samples toward a desired property, represented by a conditioning variable $y$. The guided sampling process modifies the update rule for $\rvx_t$ by incorporating a classifier score $\nabla_{\rvx_t} \log p(\rvy | \rvx_t)$ into the model's learned score function based on the relation in Eq.~\ref{eq:conditional_score}. Following \citet{song2020score}, the classifier guidance term modifies the predicted $\hat{\rvx}_0$ in the denoising process:
\begin{equation}
    \hat{\rvx}_0 = \rvx_t + \sigma^2 (s_\theta(\rvx_t, t) + \nabla_{\rvx_t} \log p(\rvy | \rvx_t)).
\end{equation}
Since our diffusion model directly predicts logits rather than the score function $s_\theta(\rvx_t, t)$, adding classifier guidance requires modifying the predicted $\hat{\rvx}_0$.

Instead of predicting the score function explicitly, our model predicts logits over the vocabulary, from which the denoised representation $\hat{\rvx}_0$ is obtained. We modify $\hat{\rvx}_0$ by incorporating classifier gradients as follows:

\begin{itemize}
    \item Compute the unmodified $\tilde{\rvx}_0$ using the model's predicted logits:
   \begin{equation}
   \tilde{\rvx}_0 = \sum_{\hat{\rvw}} p(\hat{\rvw} | \rvx_t, t) U_\theta \hat{\rvw}
   \end{equation}
   where $U_\theta$ is the embedding matrix mapping discrete tokens to continuous space.
   \item If a time-dependent classifier $f$ is available, compute the classifier guidance term:
   \begin{equation}
   \nabla_{\rvx_t} \log p(\rvy | \rvx_t) =  \nabla_{\rvx_t} f(\rvx_t, t)/\beta.
   \end{equation}
   \item Modify $\tilde{\rvx}_0$ using the classifier gradient:
   \begin{equation}
   \hat{\rvx}_0 = \tilde{\rvx}_0 + \sigma^2 \nabla_{\rvx_t} \log p(\rvy | \rvx_t).
   \end{equation}
\end{itemize}

This allows the diffusion model to generate samples that are more likely to satisfy the desired condition $\rvy$.

Further details on training the classifier are provided in Table \ref{table:classifier_training} and Table \ref{table:guidance_parameters}.

\paragraph{Discrete guidance.} \citet{nisonoff_unlocking_2024} extend classifier guidance to discrete state-space diffusion models. In analogy to classifier guidance for continuous diffusion models, they modify the unconditional rate matrix $\bar {\bm{Q}}_t$ (as defined in Eq.~\ref{eq:reverse_ctmc}) to be a conditional rate matrix $\bm R_t^\rvy$ with
\begin{equation}
    \bm R_t^\rvy[\rvx,\tilde \rvx] = \frac{p(\rvy|\tilde \rvx,t)}{p(\rvy|\rvx,t)} \bar{\bm Q}_t[\rvx,\tilde \rvx], \ \forall \tilde \rvx\neq \rvx.
\end{equation}

For classifier guidance on both continuous and discrete diffusion models, we train a time-dependent predictor (classifier) $f$ that predicts the fitness $\rvy$ given $\rvx_t$ at time $t$. We define $p(\rvy|\rvx) \propto \exp(f(\rvx)/\beta)$, where $f(\cdot)$ is a surrogate predictor of the fitness, and $\beta$ is the guidance temperature and governs the strength of guidance. Therefore, $\nabla_{\rvx_t}\log p_t(\rvy|\rvx_t) = \frac{1}{\beta} \nabla_{\rvx_t} f(\rvx_t,t)$, and $\bm R_t^\rvy[\rvx,\tilde \rvx] = \exp\left(\frac{1}{\beta}\big(f(\tilde \rvx,t) - f(\tilde \rvx, t)\big)\right) \bar{\bm Q}_t[\rvx,\tilde \rvx]$.

To obtain a classifier $f$ for discrete diffusion models, we trained an MLP regressor to predict the fitness of a one-hot encoded sequence given $\rvx_t$ and uniformly random time $t \in [0,T]$. Further details are provided in Table \ref{table:classifier_training} and Table \ref{table:guidance_parameters}.

\subsubsection{Posterior sampling}
Another line of guidance work \citep{chung2023diffusion,mardani2024a,zhang_improving_2024} focuses on drawing samples from the posterior distribution $p(\rvx|\rvy)\propto p(\rvx)p(\rvy|\rvx)$, where the prior distribution is modeled by a pretrained diffusion model. The conditional distribution $p(\rvy|\rvx)$ can either be the likelihood function of a forward model (i.e., when $\rvy$ is an incomplete measurement of $\rvx$) or an exponential distribution with respect to a reward function (i.e., $p(\rvy|\rvx) \propto \exp(f(\rvx)/\beta)$). The major difference between posterior sampling and classifier guidance is that it requires the reward function to be trained only on clean data $\rvx$. 

While many works have studied posterior sampling in Euclidean space with continuous diffusion models, posterior sampling for discrete data has been less explored. We modified DAPS \citep{zhang_improving_2024} to enable diffusion posterior sampling in discrete-state spaces. Suppose $\rvx$ lies in a finite support $\mathcal X^D$, we follow the following steps:

\begin{itemize}
    \item Initialize $\rvx_T \sim p_T(\rvx_T)$
    \item for $i = 1,\dots, K$
    \begin{itemize}
        \item[1.] Sample $\hat \rvx_0^{(i)} \sim p(\rvx_0|\rvx_{t_{i-1}})$ by a discrete diffusion model.
        \item [2.] Run Metropolis Hastings to sample $\rvx_0^{(i)} \sim p(\rvx_0| \rvx_{t_{i-1}},\rvy)$ as defined in Eq.~\ref{eq:daps_mh}.
        \item [3.] Sample $\rvx_{t_{i}} \sim p(\rvx_{t_{i}}|\rvx_0)$ following the forward Markov process.
    \end{itemize}
    \item Return $\rvx_K$.
\end{itemize}
Specifically, $t_0, t_1,\dots, t_K$ are mono-decreasing time steps with $t_0 = T$ and $t_K \approx 0$. $p(\rvx_0|\rvx_t,\rvy)$ is defined as
\begin{align}
    p(\rvx_0|\rvx_t,\rvy) &\propto p(\rvy|\rvx_0)p(\rvx_0|\rvx_t) \nonumber \\
    &\approx p(\rvy|\rvx_0) \exp(-\|\rvx_0 - \hat \rvx_0(\rvx_t)\|_0/\sigma_t),\label{eq:daps_mh}
\end{align}
where $\hat \rvx_0(\rvx_t) \sim p(\rvx_0|\rvx_t)$ is a point estimate of the conditional distribution, and we approximate $p(\rvx_0|\rvx_t)$ by an exponential distribution over Hamming distance. Following Proposition 1 in~\cite{zhang_improving_2024}, $\hat \rvx_0^{(i)}$, $\rvx_0^{(i)}$, and $\rvx_{t_i}$ converge to the posterior distribution as $t_i$ goes to 0.

For posterior sampling with DAPS, we obtained the value function $f$ using the same model architecture and training parameters as classifier guidance but only trained on clean data $\rvx$ (no noisy $\rvx_t$). We set $K = 50$ using the time scheduler for the original model. Further details are provided in Table \ref{table:classifier_training} and Table \ref{table:guidance_parameters}.

\subsubsection{NOS}

Diffusion optimized sampling (NOS)~\citep{gruver_protein_2023} is a guidance method for both continuous and discrete diffusion models, which utilizes gradient information of the continuous latent representations of protein sequences. In pretrained discrete diffusion models, noisy sequences $\rvw_t$ always have a continuous embedding in the form of hidden states of the neural network. Specifically, the denoising model that predicts $\rvw_0$ from $\rvw_t$ can be written as $p_\theta(\rvw_0|g(\rvw_t),t)$, where $\rvh_t = g(\rvw_t)$ is a continuous hidden states of the model.

Instead of training a value function on discrete sequences $\rvw_t$, NOS proposes to train the value function on the hidden states $\rvh_t$. In each diffusion step, NOS samples from the posterior distribution,
\begin{equation}
    p(\rvw_0|\rvh_t, \rvy) \propto p_\theta(\rvw_0|\rvh_t) \exp(f(\rvh_t)).
\end{equation}

To sample from this distribution, NOS runs Langevin dynamics on $\rvh_t$, i.e.,
\begin{equation}
    \rvh_t^\prime \leftarrow \rvh_t^\prime - \eta \nabla_{\rvh_t^\prime} (\lambda D_{KL}(p_\theta(\rvw_0|\rvh_t^\prime) \| p_\theta(\rvw_0|\rvh_t)) - f(\rvh_t) ) + \sqrt{2\eta\tau} \epsilon, \ \epsilon \sim \mathcal N(0,I).
\end{equation}
After $K$ iterations, we denoise $\rvw_t$ following the guided hidden state, i.e., $p(\rvw_{t-1}|\rvw_t,\rvy) = p_\theta(\rvw_{t-1}|\rvh_t^\prime, t)$.

To train the value function used for guidance in NOS, following the method from \citet{gruver_protein_2023}, we trained a very shallow neural network on the final layer hidden embeddings of the diffusion model. Further details are provided in Table \ref{table:classifier_training} and Table \ref{table:guidance_parameters}.

\subsubsection{Direct preference optimization}
For DPO with language models, we used the weighted loss function from \citet{widatalla_aligning_2024} and \citet{stocco_guiding_nodate} (Eq. \ref{eq:dpo_loss}). $\pi_\theta$ is the policy to be updated, $\pi_\text{ref}$ is the original model, and $\beta$ is a tunable parameter describing the extent of drift from the reference model. The loss therefore describes the cross entropy of the ratio $\beta \log \frac{\pi_\theta(\rvx)}{\pi_\text{ref}(\rvx)}$ and the fitness value  $w$. Following \citet{stocco_guiding_nodate}, we calculated the ratio $r$ as the difference of the log likelihood of the sequence from the updated model minus the log likelihood of the reference model, and softmax was applied to all of the fitness values $w$. We used the default parameters from \citep{stocco_guiding_nodate} and tested increasing the learning rate to $10^{-4}$ but found that generation quality broke down above the levels used in Table \ref{table:guidance_parameters} with finetuning for 5 epochs.  We also tested ranked loss with other types of models, but the performance was similar.

\begin{equation}
L_{\text{DPO}_\text{weighted}}(\pi_\theta; \pi_{\text{ref}}) = -\mathbb{E}_{D} \sum_{k=1}^{K} w^k \left[ 
    \beta \log \frac{\pi_\theta(x)}{\pi_{\text{ref}}(x)} 
    - \log \sum_{j=k}^{K} \exp\left( \beta \log \frac{\pi_\theta(x)}{\pi_{\text{ref}}(x)} \right)
\right]
\label{eq:dpo_loss}
\end{equation}

\subsection{Adaptive optimization algorithm}
\label{section:adaptive_optimization}

\begin{algorithm}[h]
\caption{Adaptive Optimization with Guided Generative Models}
\begin{algorithmic}[1]
\State \textbf{Input:} Pretrained generative prior $p(\rvx)$, initial empty labeled dataset $\mathcal{D}_0 = \emptyset$, number of rounds $T$, batch size $B$, ensemble size M
\For{$t = 1$ to $T$}
    \State Initialize batch $\mathcal{X}_t \gets \emptyset$
    \State \textbf{if} $t>1$ \textbf{then} Train ensemble of value functions $\{f_{\theta_{t,m}}\}_{m=1}^{M}$ on $\mathcal{D}_{t-1}$
        \While{$|\mathcal{X}_t| < B$}
            \If{$t>1$} 
                \State Sample value function $f_\theta \sim \text{Uniform}(\{f_{\theta_{t,m}}\}_{m=1}^{M})$ \Comment{Thompson-style sampling}
                \State Sample sequence $\rvx_b \sim \text{GuidedSample}(p(\rvx), f_\theta, \text{GuidanceStrategy})$
            \Else
                \State Sample sequence $\rvx_b \sim  \text{UnconditionalSample}(p(\rvx))$
            \EndIf
            \State \textbf{if} $\rx_b \notin \mathcal{D}_{t-1}$ \textbf{then} Add $\rvx_b$ to batch $\mathcal{X}_t$
        \EndWhile
    \State Evaluate true fitness $y_b = f_{\text{true}}(\rvx_b)$ for all $\rvx_b \in \mathcal{X}_t$
    \State Update dataset: $\mathcal{D}_{t} \gets \mathcal{D}_{t-1} \cup \{(\rvx_b, y_b)\}_{b=1}^{B}$
\EndFor
\State \textbf{Return:} Best observed sequence in $\mathcal{D}_T$
\end{algorithmic}
\end{algorithm}

We used an ensemble size of $M=10$ models, each trained with a different random initialization of neural network weights. In practice, to speed up sampling, we sampled ($B=100$ samples)/($M=10$ models) = 10 sequences in each GPU batch using the same Thompson-sampled value function, rather than using a GPU batch size of 1. Alternatively, for the Gaussian process model, we trained the model with the radial basis function kernel, and we sub-sampled the total amount of training pairs (when using noisy samples) to 5000 samples.

\subsection{Latent space Bayesian optimization with }
We utilized the APEXGo codebase \citep{torres_generative_2024}, a package for training generative variational autoencoders over peptide sequences and then optimizing those sequences with latent space Bayesian optimization to maximize certain properties. We used the training code out-of-the-box to train variational autoencoders over the same MSA sequences used to train priors for discrete diffusion models in SGPO. We trained until losses plateaued, to 542, 241, and 391 epochs for TrpB, CreiLOV, and GB1, respectively. Overall, the reconstruction losses were low, and generated sequences had low perplexity and high fitness, comparable to the generative models used in SGPO. We then used this latent space and the APEXGo optimization algorithm to maximize the fitness of sequences as measured by the oracle used in SGPO benchmarking. Specifically, in our configuration, we set the number of initialization points to 100, the number of desired diverse solutions to 1, the max number of oracle calls to 800, and the batch size to 100--to mimic the iterative Bayesian optimization experiments performed in Fig. \ref{fig:iterative}.

\subsection{Additional results}

\begin{figure}[h]
  \centering
\includegraphics[width=\textwidth]{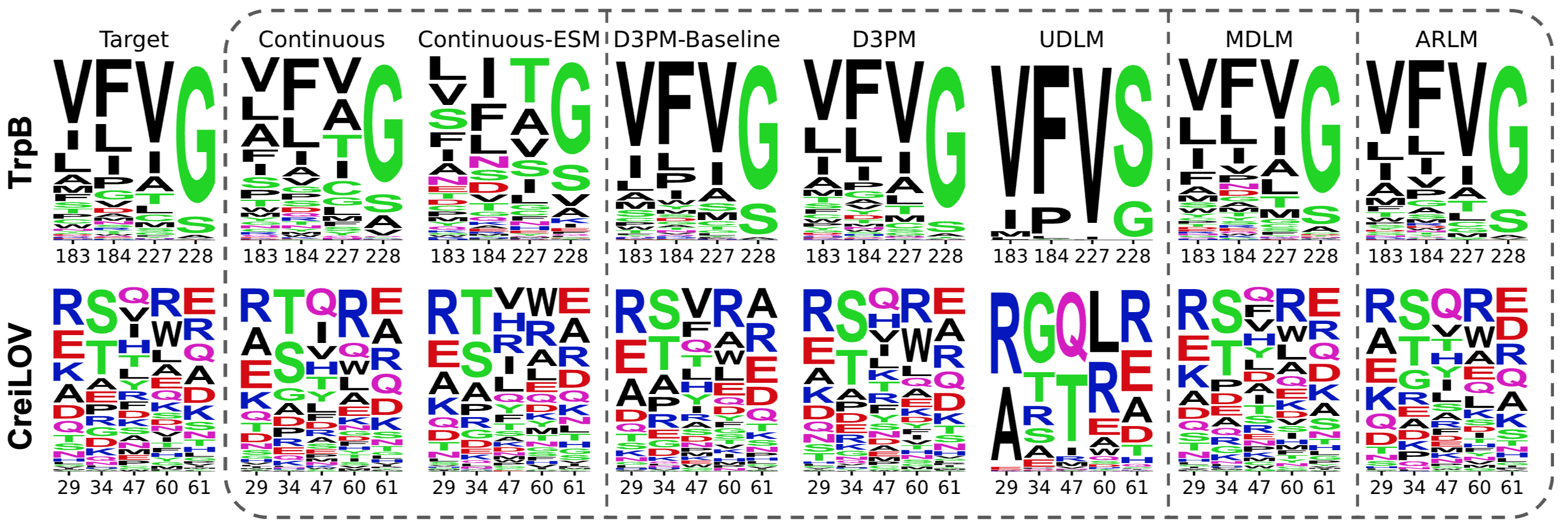}
  \vspace{-0.3in}
  \caption{\textbf{The distributions of sequences sampled from pretrained generative priors largely match those of the target distribution.} The target distribution shows all sequences in the MSA, and the distributions of generative models are approximated by sampling 1000 sequences. Model definitions can be found in Table \ref{table:prior_training}. The residues shown for TrpB are 4 out of 15 positions studied in the dataset (parent is VFVS), and 5 out of 119 residues for CreiLOV are shown as they correspond to those harboring favorable mutations in the original dataset (parent is AGQRD). Note that the target distribution for training the ARLM is slightly different than that shown here.} 
    \vspace{-0.1in}
  \label{fig:logoplot}
\end{figure}

\begin{figure}[h]
  \centering
\includegraphics[width=\textwidth]{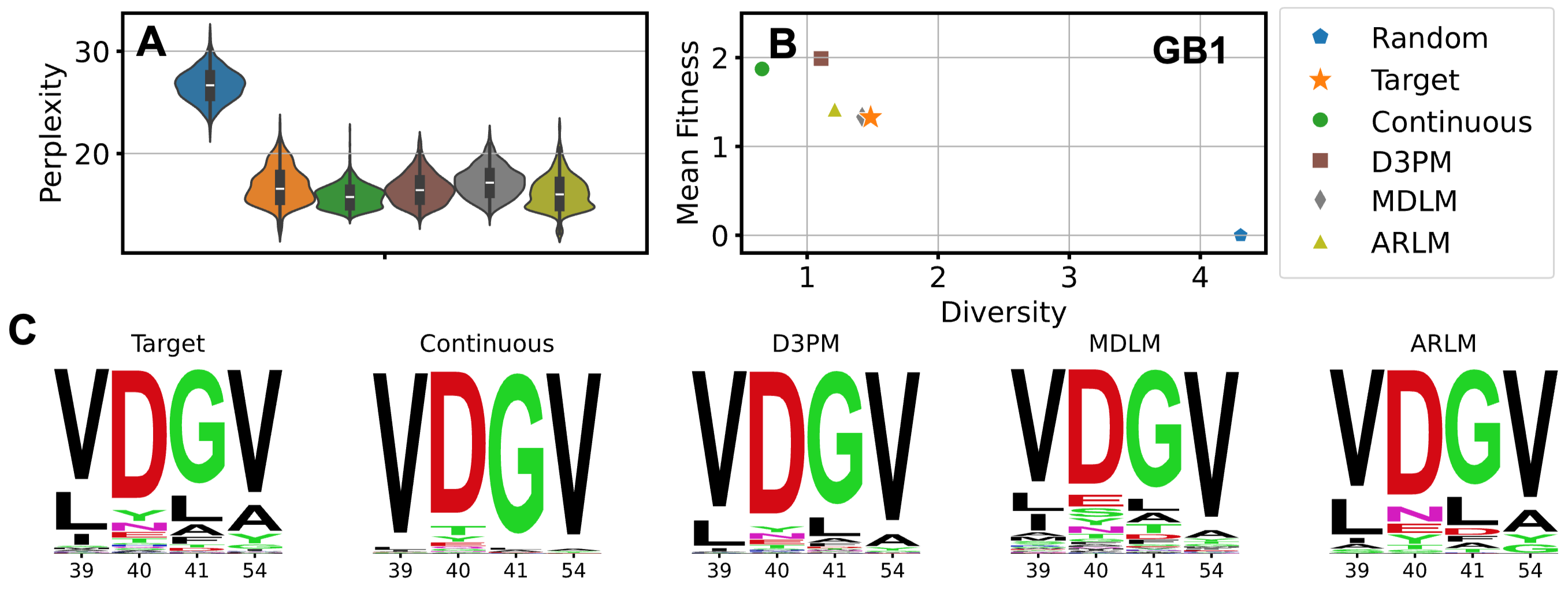}
  \vspace{-0.3in}
  \caption{Additional results for GB1, corresponding to Fig. \ref{fig:perplexity} and Fig. \ref{fig:logoplot}. The 4 positions shown correspond to the parent sequence of VDGV.}
    \vspace{-0.1in}
  \label{fig:GB1_perplexity_logoplot}
\end{figure}

\begin{figure}[h]
  \centering
\includegraphics[width=\textwidth]{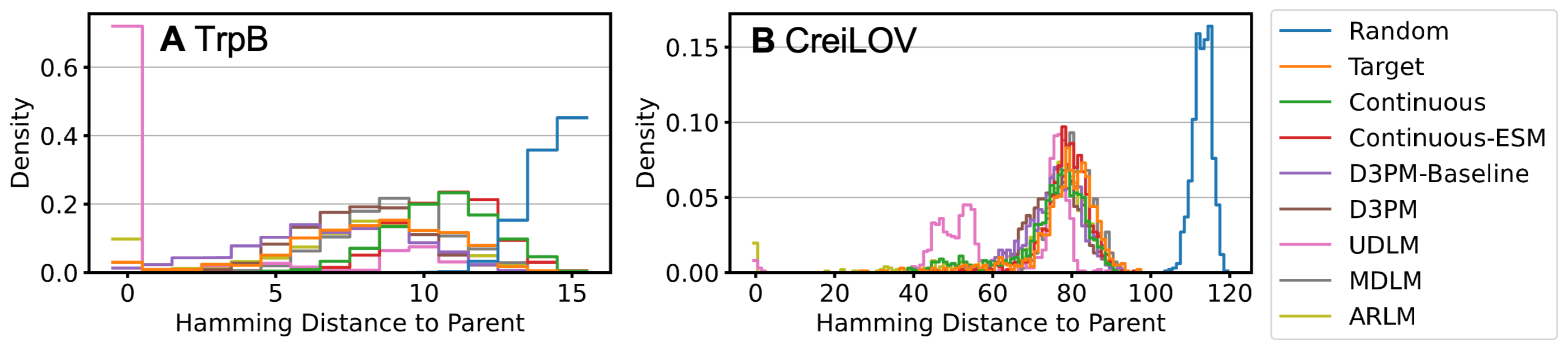}
  \vspace{-0.3in}
  \caption{Generated sequences from pretrained priors are more similar to parent than random for (\textbf{A}) TrpB and (\textbf{B}) CreiLOV, measured by the Hamming (or edit) distance. UDLM models exhibit mode collapse onto consensus sequence(s) in the training distribution. The parent sequence refers to the starting sequence used to generate variants in the original protein fitness dataset.} 
    \vspace{-0.1in}
  \label{fig:hamming_distance}
\end{figure}

\begin{figure}[h]
  \centering
\includegraphics[width=\textwidth]{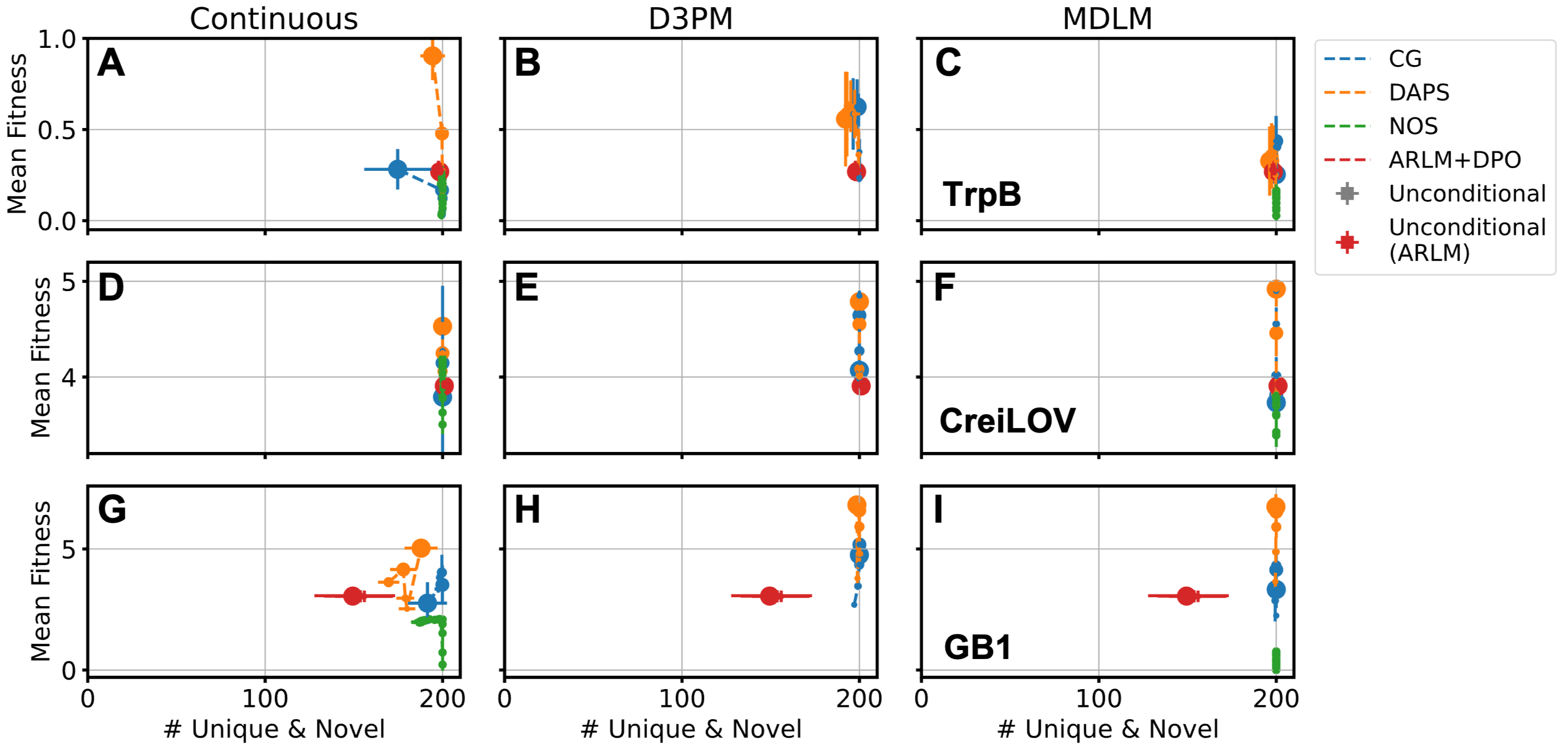}
  \vspace{-0.2in}
  \caption{Pareto boundaries demonstrate the trade-off between generating sequences with high fitness and high diversity for TrpB (\textbf{A-C}), CreiLOV (\textbf{D-F}), and GB1 (\textbf{G-I}) -- showing the same experiment as Fig. \ref{fig:fitness}. Error bars show standard deviation. Mean fitness and diversity were calculated based on 200 generated samples, with diversity calculated as the total number of unique and novel (previously unseen) samples in the generated batch, out of 200. Larger circles indicate a stronger guidance strength, specified in Table. \ref{table:guidance_parameters}.}
  \label{fig:pareto_unique}
\end{figure}

\begin{table}[h]
  \caption{Adaptive optimization with an ensemble of 10 value functions and Thompson sampling, compared to using a single model for guidance. Max fitness refers to the mean max fitness achieved at the end of the campaign using the same experimental setup as Fig. \ref{fig:iterative}, over 5 different random initializations.}
  \vspace{-0.1in}
  \centering
  \begin{tabular}{p{1.4cm}  p{1.4cm} p{1.4cm} p{2cm} p{2.2cm}} 
    \\
    \toprule
    Protein & Model  & Guidance & Max Fitness \newline (Ensemble) & Max Fitness \newline (Single Model) \\
    \midrule
    TrpB & D3PM & CG &\textbf{1.551} & 1.542 \\
    & & DAPS &\textbf{1.595} & 1.568 \\
    & MDLM & CG & \textbf{1.551} & 1.542 \\
    & & DAPS & \textbf{1.595} & 1.568 \\
    \midrule
    CreiLOV & D3PM & CG & \textbf{5.608} &  5.552 \\
    & & DAPS & \textbf{5.522} & 5.520\\
    & MDLM & CG & \textbf{5.608} & 5.552 \\
    & & DAPS & \textbf{5.530} & 5.520 \\
    \bottomrule
  \end{tabular}
  \label{table:ensemble_vs_single}
\end{table}

\begin{figure}[h]
  \centering
\includegraphics[width=11cm]{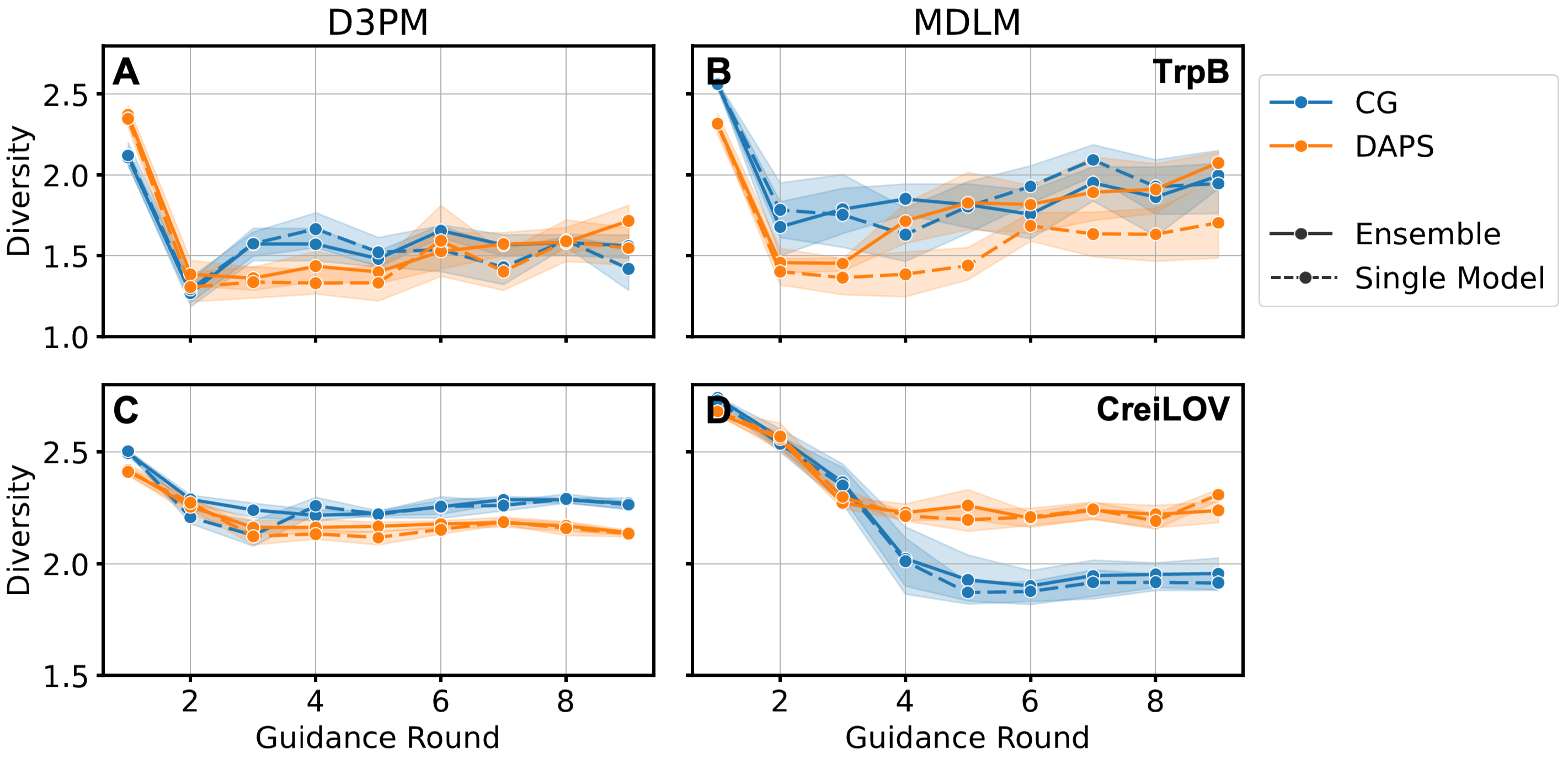}
  \vspace{-0.1in}
  \caption{Diversity of generated sequences, measured by average Shannon entropy of mutated positions, during each round of guidance. Using an ensemble of value functions and Thompson sampling generally shows higher diversity than using a single model. Experimental setup is the same as Fig. \ref{fig:iterative}, and experiments were repeated over 5 random initializations.}
  \label{fig:diversity_ensemble}
\end{figure}

\end{document}